\documentclass[11pt]{article}
\pdfoutput=1
\usepackage{jheppub,amsmath,amssymb,amsfonts,hyperref}

\usepackage{bm}
\usepackage[table]{xcolor}
\usepackage{amsmath}
\usepackage{amssymb}
\usepackage{bbold}
\usepackage{pgfplots}
\usepackage{wasysym}

\usepackage{graphicx}
\usepackage{todonotes}
\usepackage{color}
\usepackage{epsfig}

\usepackage[utf8]{inputenc}

\numberwithin{equation}{section}% numera le equazioni seconde le sezioni , e.g. 1.15 invece che consecutivamente; anche le appendici, eq.~(A.1) etc. Richiede amsmath

\definecolor{MyBlue}{rgb}{0.15,0.15,0.70}

\hypersetup{
colorlinks=true,
citecolor=MyBlue,
linkcolor=MyBlue,
urlcolor=MyBlue
}

\usepackage{amssymb}
\usepackage{amsmath}
\usepackage{amsfonts}
\usepackage{upgreek}
\usepackage{latexsym}

\newcommand{\HH}{\mathcal{H}}
\newcommand{\nn}{\nonumber}
\newcommand{\al}{\alpha}
\newcommand{\mn}{\mu\nu}

\newcommand{\lc}{\ell}

\newcommand{\fc}{f}
\newcommand{\gc}{g}
\newcommand{\MM}{\mathcal{M}}
\newcommand{\EE}{\mathcal{E}}

\newcommand{\St}{{St\"uckelberg}}
\newcommand{\Sting}{{St\"uckelberging}}

\newcommand{\pa}{\partial}

\def\ni{\noindent}

\def\be{\begin{equation}}
\def\ee{\end{equation}}
\def\bea{\begin{eqnarray}}
\def\eea{\end{eqnarray}}
\newcommand{\comment}[1]{}

\graphicspath{ {./Graphs/} }
\begin{document}

\title{On scale-free extensions of massive (bi-)gravity}
%\title{A scale-free model of bigravity}

\author[a]{Giulia Cusin,}
\author[b,c]{Nima Khosravi,}
\author[d]{Johannes Noller}

\affiliation[a]{D\'epartement de Physique Th\'eorique and Center for Astroparticle Physics,  
Universit\'e de Gen\`eve, 24 quai Ansermet, CH--1211 Gen\`eve 4, Switzerland}
\affiliation[b]{Department of Physics, Shahid Beheshti University, G.C., Evin, Tehran 19839, Iran}
\affiliation[c]{School of Physics, Institute for Research in Fundamental Sciences (IPM), P.O.Box 19395-5531, Tehran, Iran}
\affiliation[d]{Astrophysics, University of Oxford, DWB, Keble Road, Oxford, OX1 3RH, UK}

\emailAdd{giulia.cusin@unige.ch}
\emailAdd{n-khosravi@sbu.ac.ir}
\emailAdd{noller@physics.ox.ac.uk}

\date{today}
\abstract{We discuss a scale-free model of bigravity, in which the mass parameter of the standard bigravity potential is promoted to a dynamical scalar field. 
This modification retains the ghost-free bigravity structure, in particular it remains free of the Boulware-Deser ghost. 
We investigate the theory's interaction structure, focusing on its consistent scaling limits and strong coupling scales.
Furthermore we explore the model's quadratic action, both around generic background configurations and paying special attention to cosmological backgrounds and to the associated background evolution. 
Finally we consider the possibility of realizing a phase of late-time acceleration as well as a quasi-de\,Sitter inflationary stage at early times, when the promoted ``mass scalar'' becomes the inflaton.% or facilitates a late-time de Sitter phase respectively.
}


%\notoc
\setcounter{tocdepth}{2}

\maketitle
%\newpage

%\setcounter{tocdepth}{1}
%\tableofcontents

%\newpage

\section{Introduction}

The discovery of healthy massive and bi-gravity models (\cite{deRham:2010ik, deRham:2010kj, Hassan:2011vm} and \cite{Hassan:2011hr, Hassan:2011zd} respectively) paved the way for the exciting possibility that late-time cosmic acceleration could be driven by massive spin-2 degrees of freedom. However, such theories generically come with a very low strong coupling scale, which makes extracting the prediction of (a UV completion of) these theories at high energies/early times very difficult, if not impossible. For example, around Minkowski massive (bi-)gravity has a strong coupling scale $\Lambda_3 = (m^2 M_{\rm Pl})^{1/3}$, where $m$ is the (technically natural) mass of the graviton. When this mass is of order the Hubble scale -- an identification phenomenologically motivated by the desire to have this theory playing the role of dark energy at late times -- we have that $\Lambda_3 \sim 1000 km^{-1}$. While such a low strong coupling scale does not impact our ability to find cosmological solutions for this theory at late times\footnote{Note that the strong coupling scale around a cosmological background does not immediately follow from the Minkowski calculation, since this scale will be re-dressed with powers of the scale factor(s) and its derivatives. This can be worked out in detail along the lines presented in \cite{Fasiello:2013woa}.}, it seriously calls into question the predictivity of the theory at small scales (e.g. solar system tests of gravity) and at high energies (early times). 

An obvious modification of ghost-free massive and bigravity that may improve this situation is to promote the small coupling constant of the theory -- the mass of the graviton $m$ -- to a field $\Phi$ and render the theory scale-free up to the residual presence of $M_{\text{Pl}}$.\footnote{Less cosmologically relevant, but also very interesting from a theoretical perspective, would be an attempt to make the theory fully scale-free by also replacing the Planck mass(es) of the theory by scalar fields along the lines of \cite{Shaposhnikov:2008xb,Shaposhnikov:2008xi,GarciaBellido:2011de,Kannike:2015apa,Einhorn:2015lzy,Einhorn:2016mws,Ferreira:2016vsc}.} This is what we investigate in this paper. Theoretically the hope is to gain better (perturbative and weakly coupled) control of the theory via a strong coupling scale, which depends on the (local) background/vev value of the scalar field $\Phi$ and indeed we will find such a dependence. As such we will present the interaction structure of the resulting theory and show that the presence of the new dynamical scalar degree of freedom $\Phi$ leads to interesting modifications also at low energies. Cosmologically speaking the hope is that we obtain an extended bigravity model, which is capable of realising late-time accelerated expansion as well as making reliable predictions for the early universe and an associated potential inflationary period. We note that our approach is somewhat analogous to what has been done in a more generic setting for massive gravity in the context of ``mass-varying massive gravity'' models \cite{Huang:2012pe}\footnote{For related work on the cosmology of such models, see \cite{Leon:2013qh,Hinterbichler:2013dv,Gumrukcuoglu:2013nza}.}, with tight additional restrictions arising from our requirement of scale-freeness.

\textcolor{black}{
While this mass-varying nature of our model is what will allow us to have a ``running'' strong coupling scale of the theory (i.e. one that depends on the evolution of $\Phi$), this feature of course replaces the technically natural mass scale $m$ with an evolving field. As such we are faced with a mass hierarchy problem, since at late times we will require $\Phi \sim H_0$ in order for a well-behaved dark-energy contribution to arise, yet $H_0 \ll M_{\rm Pl}$.  
Scale free models can be powerful frameworks in addressing such mass hierarchy problems \cite{Shaposhnikov:2008xb,Shaposhnikov:2008xi,GarciaBellido:2011de,Kannike:2015apa,Einhorn:2015lzy,Einhorn:2016mws,Ferreira:2016vsc} and as such the hope here is that, after promoting $m$ to be a scalar field, we can dynamically generate the correct scales without the need to introduce additional scales into the action by hand. Of course, and as briefly mentioned above, our approach is only the first step in rendering the theory fully scale free, even at the classical level. Rendering the classical action fully-scale free one would need to pai our work with that of \cite{Shaposhnikov:2008xb,Shaposhnikov:2008xi,GarciaBellido:2011de,Kannike:2015apa,Einhorn:2015lzy,Einhorn:2016mws,Ferreira:2016vsc} in order to promote the residual scale $M_{\rm Pl} \to \Psi$, where $\Psi$ is a dynamical field as well and supplement this with a mechanism to dynamically recover the Planck scale, i.e. a Higgs mechanism for $\Psi$ based on spontaneous symmetry breaking that allows this field to settle to a value $\Psi \sim M_{\rm Pl}$. However, we re-emphasise that here we only tackle the first part of this problem and probe the most cosmologically interesting scale $m$ and how this can be promoted to a field in a scale-free way.\footnote{From the fully scale free perspective, this corresponds to assuming that there is a cosmologically relevant regime where spontaneous symmetry breaking has set the Planck scale already, so we can forget about the dynamics of $\Psi$ at leading order, yet $\Phi$ is still evolving.} Also note that, even classically and in spirit with  the scale-generation via spontaneous symmetry breaking argument above, around a given non-trivial cosmological background a scale will of course be generated spontaneously, as we will see explicitly in section \ref{sec-inflation}. Finally a comment on loop corrections: Once our model has been fully embedded in a fully scale-free classical theory, with all explicit mass scales in the action promoted to fields and effective scales arising via symmetry breaking, it may still be the case that loop corrections will generate an effective scale. Examples where loop corrections may break scale freeness by introducing such a mass scale include \cite{Schwinger:1962tp}, but notice that this does not always have to be the case \cite{Shaposhnikov:2008xi} and one can in principle build theories which are scale free both at classical and quantum level. Here we will focus on the first step -- constructing a scale-free classical model -- and will leave loop corrections and the stability of scale-freeness under these to future work. 
}
\\

\noindent {\it Outline}: In section \ref{sec-model} we introduce and motivate scale-free bigravity. This is followed by an investigation of its interaction structure, strong coupling scales and scaling limits in section \ref{sec-int}. Then, in section \ref{sec-Quad}, we study the general quadratic perturbative action of the theory, both for general backgrounds and specialised to cosmological Friedmann-Lema\^{\i}tre-Robertson-Walker (FLRW) backgrounds. In section \ref{sec-inflation} we finally consider specific realisations of the theory and investigate whether they give rise to accelerated expansion in the early and/or late universe, before we conclude in \ref{sec-conclusions}. We also collect some additional useful results in the appendices.
\\

\noindent {\it Notation and conventions}: Throughout we use the following conventions. We set $c=\hbar=k_{\rm Boltzmann}=1$. $M_{\text{Pl}}=1/\sqrt{8\pi G}\simeq 2.4\times 10^{18}$GeV is the reduced Planck mass. We work with the metric signature $(-, +, +, +)$, and we restrict to $D=4$ spacetime dimensions. With  $'$  and with $\cdot $ we indicate derivatives with respect to conformal time and cosmic time, respectively. %We consider only one of the two metrics coupled to matter,  and we restrict to minimal couplings. 
We use Greek letters $\mu, \nu, \ldots$ to denote spacetime indices, which are raised and lowered as specified.  Capital Latin letters $A$, $B,\ldots$ are reserved for Lorentz indices and are raised and lowered with the Minkowski metric $\eta_{AB}$.  Bracketed indices $(i),(j),\ldots$, label different fields -- label indices are not automatically summed over; whether they are upper or lower indices carries no meaning.

\section{Scale-free Bigravity} \label{sec-model}

Ghost-free bigravity \cite{Hassan:2011hr, Hassan:2011zd} typically comes with two hierarchically ordered scales: Firstly the (in principle) two Planck masses $M_{\rm Pl}^{(1)}$ and $M_{\rm Pl}^{(2)}$, directly associated to the kinetic interactions of the two spin-2 fields in the theory, which are in principle distinct but which will be identified for the purposes of this paper. As such the first scale is $M_{\rm Pl} \equiv M_{\rm Pl}^{(1)} = M_{\rm Pl}^{(2)}$. In addition there is the mass scale $m$, i.e. the parameter controlling the  mass of the massive graviton mode in the theory.\footnote{We note that in general the  mass of the graviton is a  ``dressed'' quantity into which the $\beta_i$ parameters as well as metric background quantities (if the background one expands around is dynamical), i.e. scale factors, also enter.} Note that $m$ is technically natural and protected by diffeomorphism symmetry. This ensures that the (late-)cosmologically motivated scenario with a hierarchy $m \ll M_{\rm Pl}$ can be realised, with dark energy-like self-accelerating solutions.

Here we present and discuss a very simple extension of ghost-free bigravity in which the mass scale $m$ is promoted to be a dynamical field, as motivated in the introduction above. We do this in such a way that no mass scale is introduced in the theory other than $M_{\rm Pl}$, which makes this approach different from the closely related general mass-varying massive gravity approach \cite{Huang:2012pe} (on top of the obvious fact that we also consider bi- and not massive gravity, thereby also promoting a fiducial fixed reference metric to a dynamical field - we will come back to the massive gravity limit later though).
We can write our theory down in two equivalent formulations, the vielbein and metric formulations. We will now briefly discuss both and their equivalence, since both will be useful in different contexts discussed later in the paper. 

\subsection{Vielbein and metric formulations of the theory}

\ni{\bf The vielbein version}:
The theory we propose is a very straightforward generalisation of standard ghost-free bigravity  \cite{Hassan:2011hr, Hassan:2011zd}, closely related to models of mass-varying massive gravity \cite{Huang:2012pe}. We promote the graviton mass parameter $m$ to be a scalar field of mass dimension one and endow that field with a canonical kinetic term and a potential. As such our model has the following action\small
\begin{align}
\nn {\cal S}_{\rm viel} &= \frac{M_{\rm Pl}^2}{4} \int \epsilon_{ABCD} {\bf E}^A_{(1)} \wedge {\bf E}^B_{(1)} \wedge {\bf R}^{CD}\left[E_{(1)} \right] + \frac{M_{\rm Pl}^2}{4} \int \epsilon_{ABCD} {\bf E}^A_{(2)} \wedge {\bf E}^B_{(2)} \wedge {\bf R}^{CD}\left[E_{(2)} \right]-\\
\nn &- \frac{M_{\rm Pl}^2}{2}   \epsilon_{ABCD} \int \,\Phi^2\,\Bigg( \frac{\beta_0}{4!}  \; {\bf E}^A_{(1)} \wedge {\bf E}^B_{(1)}\wedge {\bf E}^C_{(1)}\wedge {\bf E}^D_{(1)} + \frac{\beta_1}{3!}  \; {\bf E}^A_{(1)} \wedge {\bf E}^B_{(1)}\wedge {\bf E}^C_{(1)}\wedge {\bf E}^D_{(2)} +\\ \nn &+ \frac{\beta_2}{2!2!}  \; {\bf E}^A_{(1)} \wedge {\bf E}^B_{(1)}\wedge {\bf E}^C_{(2)}\wedge {\bf E}^D_{(2)} + \frac{\beta_3}{3!}  \; {\bf E}^A_{(1)} \wedge {\bf E}^B_{(2)}\wedge {\bf E}^C_{(2)}\wedge {\bf E}^D_{(2)} + \frac{\beta_4}{4!}  \; {\bf E}^A_{(2)} \wedge {\bf E}^B_{(2)}\wedge {\bf E}^C_{(2)}\wedge {\bf E}^D_{(2)} \Bigg)+ \\ 
&+ \int d^4 x \det {E_{(1)}}\Big( -\frac{1}{2} \nabla^{(1)}_\mu \Phi \nabla^\mu_{(1)} \Phi - W(\Phi) \Big) + \int d^4x \det E_{(1)} {\cal L}_m \left[E_{(1)},\Psi_i\right]\,,
\label{vielbeinAction}
\end{align}
\normalsize
formulated in terms of 2 vielbeins/spin-2 fields $E_{(1)}$ and $E_{(2)}$ and the corresponding vielbein one-forms ${\bf E}_{(i)}^A \equiv {E}_{(i)}{}_\mu^A dx^\mu$. This means each spin-2 field comes equipped with an Einstein-Hilbert term (first line), we have massive bigravity interactions with a graviton mass that has been promoted to be a field $\Phi$ (second and third line), we have an additional piece of the action giving $\Phi$ dynamics (first term in the last line) and finally we have a minimal coupling of matter, via the matter Lagrangian ${\cal L}_m$ containing the matter fields $\Psi_i$, to one of the vielbeins $E_{(1)}$ (second term in the last line).\footnote{Note that this matter coupling breaks the symmetry between the spin-2 fields/metrics/vielbeins in the theory. For consistent couplings that restore this symmetry see \cite{deRham:2014naa,Noller:2014sta,Melville:2015dba}.}  Note that there are five dimensionless  coupling constants $\beta_i$ in addition to the dimensionful coupling constant $M_{\rm Pl}$. In an effective field theory spirit we will consider the $\beta_i$ to be constant ${\cal O}(1)$ parameters. The metric corresponding to each vielbein satisfies 
\be\label{1}
g_{\mn}^{(i)} = E^A_{(i)\mu} E^B_{(i)\nu} \eta_{AB}\,,
\ee 
%Each vielbein $E_{(i)}$ has associated a metric $g_{(i)}$ and a 
and comes with an associated covariant derivative $\nabla_{(i)}$, while the wedge product $\wedge$ in \eqref{vielbeinAction} has been defined as totally anti-symmetrising space-time indices as usual. 

The potential for $\Phi$, $W(\Phi)$ is in principle unrestricted. However, if we insist that there are no other dimensionful scales in the theory other than $M_{\rm Pl}$, i.e. we forbid any dimensionful scales from hiding in $W(\Phi)$, this means we can write
\be\label{W choice}
W(\Phi) = \lambda \Phi^4\quad \text{or}\quad W(\Phi)=0\,,
\ee 
where $\lambda$ is a dimensionless parameter of arbitrary size\footnote{Such a term is renormalizable, so we do not need to insist on this being ${\cal O}(1)$ even from an EFT perspective.}. Note that the constraint structure of healthy massive and bi-gravity models carries over and ensures that no ghostly Boulware-Deser degrees of freedom propagate \cite{Huang:2012pe}. As such, around a flat Minkowski background, we have 8 propagating degrees of freedom: 5 (massive graviton) + 2 (massless graviton) + 1 (the new scalar $\Phi$), one dimensionful scale/mass parameter $M_{\rm Pl}$ and six dimensionless parameters $\beta_i, \lambda$. For a more detailed discussion regarding degree of freedom counting for analogous models see \cite{deRham:2014zqa}.
\\

%\subsubsection{The metric version}
\ni{\bf The metric version}:
For comparison we can also write down the metric version of our theory. In the presence of the symmetric vielbein condition $E_{(i)}{}^A_\mu E_{(j)}{}^B_\nu \eta_{AB} = E_{(i)}{}^A_\nu E_{(j)}{}^B_\mu \eta_{AB}$, which we discuss further below, the two versions become physically equivalent. In the metric picture our theory takes on the form 
\begin{align}
{\cal S}_{\rm metric} &= \frac{M_{\rm Pl}^2}{2} \int d^D x \sqrt{-g_{(1)}} R[g_{(1)}] + \frac{M_{\rm Pl}^2}{2} \int d^D x \sqrt{-g_{(2)}} R[g_{(2)}]  -\nn \\
&- \frac{M_{\rm Pl}^2}{2}  \sum_{n=0}^{D} \beta_n \int d^Dx  \sqrt{-g_{(1)}} \, \Phi^2\,e_n\left(\sqrt{g^{-1}_{(1)} g_{(2)}}\right)+\nn \\
&+ \int d^D x \sqrt{-g_{(1)}}\Big( -\frac{1}{2} \nabla_\mu^{(1)} \Phi  \nabla^\mu_{(1)} \Phi - W(\Phi) \Big) +\nn \\
&+ \int d^D x \sqrt{-g_{(1)}} {\cal L}_m \left[g_{(1)},\Psi\right]\,,
\label{metricAction}
\end{align}
where we will take $D=4$ in what follows and for later convenience we define a potential 
\be
V\left(g_{(1)}\,, g_{(2)}\right)\equiv \sum_{n=0}^{D} \beta_n \, e_n\left(\sqrt{g^{-1}_{(1)} g_{(2)}}\right)\,.
\ee
 The $\beta_i$ are the same as above and the $e_n$ are elementary symmetric polynomials satisfying (for some matrix $\mathbb{X}$)
\be
e_n\left(\mathbb{X}\right) = \delta^{\alpha_1 \ldots \alpha_n}_{[\beta_1 \ldots \beta_n]} \mathbb{X}^{\beta_1}_{\alpha_1} \cdots \mathbb{X}^{\beta_n}_{\alpha_n},
\ee
where we have defined 
\be \label{delta-def}
\delta^{\alpha_1 \ldots \alpha_n}_{[\beta_1 \ldots \beta_n]}  \equiv \frac{1}{n!(D-n)!}\varepsilon^{\alpha_1 \ldots \alpha_n \lambda_1 \ldots \lambda_{D-n}}
\varepsilon_{\beta_1 \ldots \beta_n \lambda_1 \ldots \lambda_{D-n}}.
\ee
As such, the elementary symmetric polynomials can explicitly be written as
\begin{eqnarray}
e_{0}(\mathbb{X}) &=& 1,\, \quad e_{1}(\mathbb{X})=[\mathbb{X}],\, \quad e_{2}(\mathbb{X}) =\frac{1}{2!}\left([\mathbb{X}]^{2}-[\mathbb{X}^{2}]\right)\,,\nn \\ 
e_{3}(\mathbb{X}) &=& \frac{1}{3!}\left([\mathbb{X}]^{3}-3[\mathbb{X}][\mathbb{X}^{2}]+2[\mathbb{X}^{3}]\right)\,,\nn \\
e_{4}(\mathbb{X}) &=& \frac{1}{4!}\left([\mathbb{X}]^{4}-6[\mathbb{X}]^{2}[\mathbb{X}^{2}]+8[\mathbb{X}][\mathbb{X}^{3}] +3[\mathbb{X}^{2}]^{2}-6[\mathbb{X}^{4}]\right) = \det \mathbb{X} \,,
\end{eqnarray}
%where we have defined the shorthand $\mathbb{X} \equiv \sqrt{g^{-1}_{(1)} g_{(2)}}$, 
where square brackets $\left[\cdots\right]$ denote taking the trace. 
%and the numerical prefactors are analogous to those for each $\beta_i$ in the vielbein action. 
While we will use the metric formulation of the theory when computing cosmological solutions later, we will find the vielbein formulation more useful when probing the interaction structure of the theory.
\\

%\subsubsection{Equivalence of formulations}
\ni{\bf Equivalence of formulations}:
Above we have discussed both the vielbein and metric formulations of our model. Working in different formulations will be useful in what follows below, but we want to briefly recap why these two formulations are equivalent in our context. On certain branches of solutions (the ones we will consider - for a more complete discussion including alternative branches see \cite{Deffayet:2012zc,Banados:2013fda}) the ``symmetric vielbein condition'' can be enforced. This is equivalent to the statement that we can set the so-called DvN (Deser-van Nieuwenhuizen) gauge\footnote{Essentially this gauge choice fixes the freedom associated with Lorentz transformations.} which imposes the following relation between two distinct vielbeins $E_{(i)}$ and $E_{(j)}$ (in matrix notation)
\be
E_{(i)}^{-1} E_{(j)} \eta = \eta \left(E_{(i)}^{-1} E_{(j)}\right)^T\,,
\ee
where $\eta$ denotes the flat Minkowski metric as before. We can then use this condition and the expressions of the metrics in terms of their vielbeins to find \cite{Hinterbichler:2012cn}
\be
\int d^4x \sqrt{-g_{(i)}}\sum_n \beta_n e_n\left(\sqrt{g_{(i)}^{-1}g_{(j)}} \right) = 
\int d^4x \det E_{(i)} \sum_n \beta_n e_n\left(E_{(i)}^{-1} E_{(j)} \right),
\ee
which relates our two formulations and identifies the two actions \eqref{vielbeinAction} and \eqref{metricAction} (and in particular the mass terms and the $\beta_i$ coefficients) as equivalent upon noticing that we can re-write the mass terms in the following way

\begin{align}
& &&\frac{M_{\rm Pl}^2}{2}  \epsilon_{ABCD} \int \frac{\beta_3}{3!(D-3)!}   \Phi^2\,\; {\bf E}^A_{(1)} \wedge {\bf E}^B_{(2)}\wedge {\bf E}^C_{(2)}\wedge {\bf E}^D_{(2)} \nn \\
=& &&\frac{M_{\rm Pl}^2}{2} \frac{\beta_{3}}{3!(D-3)!} \int d^4x \;  \Phi^2\,\epsilon_{ABCD}\epsilon^{\mu\nu\rho\sigma} \;  {E}^A_{\mu}{}_{(1)} {E}^B_{\nu}{}_{(2)} {E}^C_{\rho}{}_{(2)} { E}^D_{\sigma}{}_{(2)}  \nn \\ 
=& &&\frac{M_{\rm Pl}^2}{2} \beta_3 \int d^4x\,  \Phi^2\,\det E_{(1)} e_3\left(E_{(1)}^{-1} E_{(2)} \right)\nn \\ 
=& &&\frac{M_{\rm Pl}^2}{2}  \beta_3 \int d^4x \, \Phi^2\,\sqrt{-g_{(1)}} e_3\left(\sqrt{g_{(1)}^{-1}g_{(2)}} \right)\,,
\end{align}
where we have picked one particular mass term and $\beta_i$ for illustration, but have kept all numerical arguments as explicit as possible to emphasize that an analogous argument follows through for all other mass terms too. Notice how all residual factorial factors are swallowed up into $e_3$ in moving to the third line. In moving from the second to the third line we have also extracted an overall factor of $\det E_{(1)}$, where this choice is arbitrary (i.e. we could just as well have extracted $\det E_{(2)}$). %and fields $1 \leftrightarrow 2$ could be reversed without any physical effect.

\subsection{Equations of motion and constraints}

\ni{\bf Equations of motion}:
We now move on to consider the dynamics of our model at the level of the equations of motion. We will do so in the metric language and, in order to keep notation clean, will define $g_{(1)} = g$ and $g_{(2)} = f$. The equations of motions for $g_{\mu\nu}$ and $f_{\mu\nu}$ are then given by
\begin{align}
R_{\mu\nu}(g)-\frac{1}{2}g_{\mu\nu}\,R(g)-\Phi^2\,\mathcal{M}_{\mu\nu}^{g}&=\frac{1}{M_{\text{Pl}}^2}\, T_{\mu\nu}^{(m)}+\frac{1}{M_{\text{Pl}}^2}\mathcal{T}_{\mu\nu}^{\Phi}\,,\\
R_{\mu\nu}(f)-\frac{1}{2}f_{\mu\nu}\,R(f)-\Phi^2\, \mathcal{M}_{\mu\nu}^{f}&=0\,,
\end{align}
where to simplify the notation we have introduced the following tensors\small
\begin{align}
&\mathcal{M}_{\mu\nu}^{g}\equiv \frac{\delta V}{\delta g^{\mu\nu}}-\frac{1}{2}g_{\mu\nu}\,V=\frac{1}{4}\, \sum_{n=0}^3(-)^{n+1}\, \beta_n\, \left[g_{\mu\lambda}\, Y_{(n)\nu}^{\lambda}\left(\sqrt{g^{-1}f}\right)+g_{\nu\lambda}\, Y_{(n)\mu}^{\lambda}\left(\sqrt{g^{-1}f}\right)\right]\,,\\
&\mathcal{M}_{\mu\nu}^{f}\equiv \frac{\delta V}{\delta f^{\mu\nu}}-\frac{1}{2}f_{\mu\nu}\,V= \frac{1}{4}\, \sum_{n=0}^3(-)^{n+1}\, \beta_{4-n}\,\left[ f_{\mu\lambda}\, Y_{(n)\nu}^{\lambda}\left(\sqrt{f^{-1}g}\right)+f_{\nu\lambda}\, Y_{(n)\mu}^{\lambda}\left(\sqrt{f^{-1}g}\right)\right]\,,
\end{align}
\normalsize
while $T_{\mu\nu}^{(m)}$ is the energy momentum tensor of matter and 
\begin{align}\label{tau}
\mathcal{T}_{\mu\nu}^{\Phi}\equiv\nabla_{\mu}\Phi\,\nabla_{\nu}\Phi-g_{\mu\nu}\left(\frac{1}{2}\,g^{\alpha\beta}\nabla_{\alpha}\Phi\,\nabla_{\beta}\Phi+W(\Phi)\right)\,.
\end{align}
The definition of the $Y_{(n)\mu}^{\nu}\left(\mathbb{X}\right)$ matrices closely mimics that of the elementary symmetric polynomials and is as follows:
\begin{align}
&Y_{(0)}\left(\mathbb{X}\right)=\mathbb{I}\,,\hspace{0.5 cm} Y_{(1)}\left(\mathbb{X}\right)=\mathbb{X} -\mathbb{I}\left[\mathbb{X}\right]\,,\\
&Y_{(2)}\left(\mathbb{X}\right)=\mathbb{X}^2 -\mathbb{X}\left[\mathbb{X}\right]+\frac{1}{2}\mathbb{I}\left(\left[\mathbb{X}\right]^2-\left[\mathbb{X}^2\right]\right)\,,\\
&Y_{(3)}\left(\mathbb{X}\right)=\mathbb{X}^3 -\mathbb{X}^2\left[\mathbb{X}\right]+\frac{1}{2}\mathbb{X}\left(\left[\mathbb{X}\right]^2-\left[\mathbb{X}^2\right]\right)-\frac{1}{6}\mathbb{I}\left(\left[\mathbb{X}\right]^3-3\left[\mathbb{X}\right]\left[\mathbb{X}^2\right]+2\left[\mathbb{X}^3\right]\right)\, .
\end{align}
The equation of motion for the field $\Phi$ can be written as
\be
\Box \Phi-W_{,\Phi}(\Phi)- M_{\text{Pl}}^2\, \Phi \,V\left(g, f\right)=0\,,
\ee
where $W_{,\Phi}\equiv d W/d \Phi$. 
\\

\ni{\bf Bianchi constraints}:
As a consequence of the Bianchi identity, we find the following Bianchi constraints (for each one of the two metrics)
\be\label{b1}
\nabla^{\mu}\left(\Phi^2\,\mathcal{M}_{\mu\nu}^{g}+\frac{1}{M_{\text{Pl}}^2}\mathcal{T}_{\mu\nu}^{\Phi}\right)=-\nabla^{\mu}T_{\mu\nu}^{(m)}\,,
\ee
\be\label{b2}
\bar{\nabla}^{\mu}\left(\Phi^2\,\mathcal{M}_{\mu\nu}^{f}\right)=0\,,
\ee
where the overbar indicates covariant derivatives with respect to the $f$ metric. Both these constraints follow from the invariance of the action under the diagonal subgroup of the general coordinate transformations of the two metrics. 

It is easy to show that the Bianchi  constraint (\ref{b1}) is equivalent to the covariant conservation of the total energy momentum tensor
\be
T_{\mu\nu}\equiv T_{\mu\nu}^{\Phi}+T_{\mu\nu}^{(m)}\,,
\ee
where $T_{\mu\nu}^{\Phi}$ is the energy momentum tensor for the scalar field. 
The lagrangian for $\Phi$ is
\be
\mathcal{L}^{\Phi}=-\frac{1}{2}\left(\nabla_{\mu}\Phi\right)^2-W(\Phi)-\frac{M_{\text{Pl}}^2}{2}\,\Phi^2\, V\left(g,f\right)\,.
\ee
Therefore, the energy momentum tensor for the scalar field is given by
\begin{align}\label{enmominflaton}
T_{\mu\nu}^{\Phi}&=-2\,\frac{\delta \mathcal{L}^{\Phi}}{\delta g^{\mu\nu}}+g_{\mu\nu}\,\mathcal{L}^{\Phi}\,,\nn\\
&=\nabla_{\mu}\Phi\,\nabla_{\nu}\Phi+M_{\text{Pl}}^2\,\Phi^2\,\frac{\delta V}{\delta g^{\mu\nu}}-g_{\mu\nu}\left(\frac{1}{2}\,g^{\alpha\beta}\nabla_{\alpha}\Phi\,\nabla_{\beta}\Phi+W(\Phi)+\frac{M_{\text{Pl}}^2}{2}\,\Phi^2\,V\right)\,,\nn\\
&=\mathcal{T}_{\mu\nu}^{\Phi}+M_{\text{Pl}}^2\,\Phi^2\,\mathcal{M}_{\mu\nu}^{g}\,.
\end{align}
It follows that
\be
\nabla^{\mu}T_{\mu\nu}=\nabla^{\mu}T_{\mu\nu}^{\Phi}+\nabla^{\mu}T_{\mu\nu}^{(m)}=\nabla^{\mu}\left(\mathcal{T}_{\mu\nu}^{\Phi}+M_{\text{Pl}}^2 \Phi^2\,\mathcal{M}_{\mu\nu}^{g}\right)+\nabla^{\mu} T_{\mu\nu}^{(m)}=0\,,
\ee
where the last identity follows from eq. (\ref{b1}).

\section{Interaction structure and strong coupling scales} \label{sec-int}

We now take our model \eqref{vielbeinAction} and make the propagating degrees of freedom explicit in order to better understand the interaction structure of the theory. For this it will turn out to be useful to work in the vielbein formulation. A priori one might expect that the interaction structure is very different with respect to the ``standard'' bi-gravity case, especially since scalar modes are expected to mix already at the level of the quadratic action, which would result in a different diagonalisation procedure and different dynamics for the propagating degrees of freedom. Also, several of the interactions which vanish up to total derivatives in the standard bigravity case will now remain, since the ``mass scale prefactor'' is now dynamical. % and these terms no longer all vanish up to total derivatives as a result. 
 As such we will take particular care in going through the derivation and will not simply port expressions or field normalisations from the analogous bigravity calculation \cite{Fasiello:2013woa}.
\\

\subsection{The field content}

\ni {\bf \St{} fields and degrees of freedom}: 
%\subsubsection{\St{} fields and degrees of freedom}
We want to restore diffeomorphism invariance in our action in order to make the dynamics of the different helicity modes explicit (as the helicity 0, 1 and 2 modes are all bundled together in $h_{(1)}$ and $h_{(2)}$ which are perturbations around the background for $E_{(1)}$ and $E_{(2)}$ respectively). As such let us begin by briefly recapping the use of \St{} fields to restore diffeomorphism invariance. We have two vielbeins in the theory, $E_{(1)}$ and $E_{(2)}$, transforming under two copies of general coordinate transformations, $GC_{(1)}$ and $GC_{(2)}$. The mass interaction term(s) break this invariance down to the diagonal subgroup $ GC_{(1)} \times GC_{(2)}$. The \St{} trick then amounts to restoring the full unbroken invariance at the expense of introducing additional gauge fields, which will eventually turn out to capture the different helicity degrees of freedom of the graviton(s) in the theory. In effect the diffeomorphism \St{} replacement amounts to a field transformation of one of the vielbeins
\be
E_\mu^A{}_{(2)}[x] \to \tilde E_\mu^A{}_{(2)}[Y(x)] =  E_\nu^B{}_{(2)}[Y(x)] \pa_\mu Y^\nu_{(2,1)}[x] \label{stuck}\,.
\ee
Here we have chosen to transform $E_{(2)}$, but this choice is of course arbitrary. For a discussion of dualities and ambiguities in the context of choosing \St{} transformations for bi- and multi-gravity models see \cite{Noller:2013yja,Noller:2015eda}.\footnote{The different dual ways of doing so are also directly related to the existence of ``Galileon dualities''. \cite{deRham:2013hsa}.}  The kinetic, Einstein-Hilbert, terms are gauge-invariant under diffeomorphisms, so remain unmodified under this replacement. We may now expand the vielbeins and \St{} fields as
\begin{align}
E_\mu^A{}_{(i)} &= \delta_\mu^A + \frac{h_\mu^A{}_{(i)}}{2 M_{\rm Pl}} , 
&Y^\nu_{(i,j)} &= x^\nu_{(i)} + B^\nu_{(i,j)} + \pa^\nu \pi_{(i,j)},
\label{decexp}
\end{align}
effectively choosing to expand our theory around a Minkowski background. In (\ref{decexp}), $h_{(1)}$, $h_{(2)}$, $B$, $\pi$ are the fields which will capture the two helicity-2 modes as well as helicity-1 and -0 modes of the massless and massive graviton respectively. Note that we have already chosen to canonically normalise the $h$ fields in the above, since this normalisation will be controlled by the known Einstein-Hilbert term. The normalisation of the other fields is left for later, since this will be determined by the quadratic action for those fields. In what follows for simplicity we will drop the combined $(i,j)$ indices, that keep track of the symmetry group(s) the \St{} fields know about, since having chosen to transform $E_{(2)}$ here, there will only be one $\pi$ and one $B$ field.
 
Another note is in order before we proceed: in the vielbein formulation it is not just copies of general coordinate invariance which interactions break down to their diagonal subgroup, but the same also happens for Lorentz invariance. The Lorentz \St{} fields will crucially modify the interactions of the helicity-1 mode, which is why in the following we contain ourselves to investigating the helicity-2/0 interactions and we leave an investigation of interactions involving helicity-1 modes for future research - for an analogous calculation in the standard massive (bi-)gravity setting, see \cite{Ondo:2013wka,Fasiello:2013woa}. From here on we will therefore set $B = 0$.
\\

%\todo[inline]{I have restored the use of $\phi_0$ below and tried to make things consistent again (in the last version $\phi$ was used both as a background field and as a fixed scale). I can't see any clashes with notation anywhere else right now and alternatives such as $\hat \phi, \bar \phi$ have the conotation of a background field, whereas $\phi_0$ is also what would be used as a zeroth order (constant) term in Taylor expansions for example, so it's the best notation I can think of. Let me know if you have better suggestions - the main issue here is to make it as clear as possible that the $\phi_0$ field is NOT dynamical.}

\ni {\bf A local field expansion and scaling limits}: 
%\subsubsection{A local field expansion}
%
There is an inherent tension in what we are trying to achieve here. On the one hand we have intentionally built an, except for $M_{\rm Pl}$, scale-free theory. Yet on the other hand we are here trying to obtain a perturbative understanding of the interaction structure of the theory. The helicity-0 mode will inherit its normalisation from the mass term, which now comes with a time-dependent ``mass scale'' $\phi$. From the form of the interactions one should therefore expect that the scales determining when any particular perturbative expansion of the theory is valid (strong coupling scales), will now depend on the background value of $\Phi$ (or, in a different language, on its vev). In order to make this explicit and understand the evolving strong coupling scale while simultaneously maintaining our scale-free theory, we will perform a local expansion of $\Phi$ around some fixed  reference value $\phi_0$. We emphasise that $\phi_0$ is fixed, so it is {\it not} a dynamical background field. In other words we will perform the split
\begin{align}
\Phi &= \phi_0 + \delta\phi &&\text{where} &\delta\phi &\ll \phi_0\,.
\end{align}
The condition $\delta\phi \ll \phi_0$ ensures that we can normalise modes coming from the mass term using $\phi_0$ at leading order and have a well-defined perturbative expansion in powers of $\delta\phi$. Obviously this will only give a locally valid expansion, as there is no guarantee that the evolution of $\Phi$ will not eventually lead to $\delta\phi \gtrsim \phi_0$ for an arbitrary previously chosen $\phi_0$. However, locally (by which we mean: local in space-time, but particularly ``local'' in time) this will be a useful expansion to use. In this way we will get a handle on what interactions exist locally, for which configurations the perturbation theory breaks down and what the relevant strong coupling scales are. 
\\

\subsection{Propagating degrees of freedom}

\ni {\bf Tadpole cancellations}: 
%\subsubsection{Tadpole cancellations}
Armed with the above \St{} and field expansion schemes, we can now go through the action order-by-order. Throughout we will ignore contributions coming from $W(\Phi)$ -- these can straightforwardly be added once a concrete form for this potential is specified.\footnote{Note that a power-law $W(\Phi)$ will typically introduce a tadpole for $\delta\phi$, which cannot be removed. This effectively just means that $\delta\phi = 0$ is not a solution of the theory in general in this case.}  First up are terms linear in the fields, i.e. tadpole terms. We would like to remove these terms so that the backgrounds we have chosen really are solutions of the theory\footnote{In other words we require that Minkowski is a solution, which imposes these extra conditions.}. This will impose a condition on the dimensionless order one coefficients $\beta_i$ of the theory. Using the above expansions we find that the resulting linear terms are given by
\begin{align}
\text{``tadpole terms''} \quad &= \quad  M_{\text{Pl}}^2 (\beta_0^{\text{}} + 4 \beta_1^{\text{}} + 6 \beta_2^{\text{}} + 4 \beta_3^{\text{}} + \beta_4^{\text{}}) \phi_0 \,\delta \phi + \nn \\
&+ \quad \frac{M_{\text{Pl}}}{2} (\beta_0^{\text{}} + 3 \beta_1^{\text{}} + 3 \beta_2^{\text{}} + \beta_3^{\text{}}) \phi_0^2 \,h^{\mu}{}_{\mu} + \nn \\
&+ \quad \frac{M_{\text{Pl}}}{2} (\beta_1^{\text{}} + 3 \beta_2^{\text{}} + 3 \beta_3^{\text{}} + \beta_4^{\text{}}) \phi_0^2 \, l^{\mu}{}_{\mu}\,,
\label{tadpoles}
\end{align}
where we have defined $h \equiv h_{(1)}$ and $l \equiv h_{(2)}$ to avoid clutter. Removing these terms imposes the following conditions (which we choose to express as conditions on $\beta_0$ and $\beta_4$): 
\begin{align}
\beta_0 &= -3\beta_1 - 3\beta_2 - \beta_3\,, \nn \\
\beta_4 &= -\beta_1 - 3\beta_2 - 3\beta_3\,.
\label{betasol}
\end{align}
In what follows we will impose those conditions, so that we are in effect left with three dimensionless ${\cal O}(1)$ parameters: $\beta_1,\beta_2,\beta_3$.
\\

\ni {\bf The quadratic action}: 
%\subsubsection{The quadratic action}
 Next up is the quadratic action, which importantly will determine how we have to normalise the $\pi$ field. We should expect mixing not just between the $h$ fields and $\pi$ (scalar-tensor mixing) but also between $\delta\phi$ and $\pi$ (scalar-scalar mixing), both of which should be removed via diagonalising transformations. We begin by ignoring mass terms (i.e. quadratic non-derivative interactions) and look at the kinetic interactions at the quadratic level. We split these into pure scalar, pure tensor and scalar-tensor interactions
\be
{\cal S}^{\rm kin}_2 = {\cal S}^{\rm kin}_{\rm scalar} + {\cal S}^{\rm kin}_{\rm tensor} + {\cal S}^{\rm kin}_{\rm scalar-tensor}\,.
\ee

We first look at pure scalar interactions, involving the \St{} field $\pi$ and the scalar $\delta\phi$. The field $\pi$ does not have its own kinetic term and in the standard bigravity case obtains its kinetic term via demixing from the tensors. Here scalar interactions are in principle also mixed, which (after demixing the scalars) would give rise to an apparent ghost. Explicitly we find
\be 
{\cal S}^{\rm kin}_{\rm scalar} = \int d^4x \left[-  \tfrac{1}{2} \partial_{\mu}\delta \phi \partial^{\mu}\delta \phi +
M_{\text{Pl}}^2 (\beta_1^{\text{}} + 3 \beta_2^{\text{}} + 3 \beta_3^{\text{}} + \beta_4^{\text{}}) \phi_0^{\text{}} \delta \phi \pi^\mu_\mu\right]\,,
\ee
where we have used the shorthand $\pi_\mu^\nu \equiv \pa_\mu \pa^\nu \pi$.
This immediately looks dangerous, since the associated kinetic mixing matrix has opposite sign eigenvalues, so one of the two modes would behave as a ghost. However, a closer look shows that the tadpole cancellation requirements from above in fact eliminate the scalar mixing term and as a result the scalar action simply reduces to
\be \label{SkinScalar}
{\cal S}^{\rm kin}_{\rm scalar} = \int d^4x  \left[-\tfrac{1}{2}\pa_\mu \delta\phi\pa^\mu \delta\phi\,\right]\,.
\ee
This means that $\pi$ will have to inherit its kinetic term from scalar-tensor mixing terms as usual and that $\delta\phi$ is automatically decoupled from the other fields at quadratic level. As such the rest of this section can proceed just as for the standard bigravity case. 

Moving on we now consider pure tensor and scalar terms together, where we recall that we defined $h \equiv h_{(1)}$ and $l \equiv h_{(2)}$. We focus on the $h-\pi$ mixing (the argument will be the same for $l-\pi$). Pure tensor interactions for $h$ (and analogously for $l$) are given by
\begin{align} 
{\cal S}^{\rm kin}_{\rm tensor} &= \int d^4x  \left[\tfrac{1}{8} h^{\mn} \partial_{\mu}\partial_{\nu}h^{\rho}{}_{\rho} 
-\tfrac{1}{4} h^{\mn} \partial_{\rho}\partial_{\nu}h_{\mu}{}^{\rho} + \tfrac{1}{8} h^{\rho}{}_{\rho} \partial_{\mu}\partial_{\nu}h^{\mn} + \tfrac{1}{8} h^{\mn} \Box h_{\mn} -  \tfrac{1}{8} h^{\mu}{}_{\mu} \Box h^{\nu}{}_{\nu} \right] \nn \\
&\equiv \int d^4x \; {\cal L}_2^{\rm EH}(h)\,,
\end{align}
i.e. a linearised Einstein-Hilbert term, whereas the scalar-tensor mixing interactions between the scalar $\pi$ and $h$ at quadratic order are
\begin{align}
{\cal S}^{\rm kin}_{\rm scalar-tensor} =  \tfrac{1}{2}M_{\text{Pl}} \phi_0^2 (\beta_1^{\text{}} + 2 \beta_2^{\text{}} + \beta_3^{\text{}}) \int d^4x \left[ h^{\mu}{}_{\mu} \Box\pi - h^{\mn} \pi_{\mn}\right].
\end{align}
Note that here we have already substituted in all the expressions above the expression for $\beta_0,\beta_4$ in eq. \eqref{betasol}, coming from tadpole cancellation requirements, as we will in what follows throughout this section. 
Demixing these $h-\pi$ interactions and the analogous $l-\pi$ interactions amounts to performing the following two linearised conformal transformations
\begin{align}
h_{\mu\nu} &\to h_{\mu\nu} +  M_{\text{Pl}} \phi_0^2 \left( \beta_1^{\text{}} + 2  \beta_2^{\text{}} + \beta_3^{\text{}}  \right)\eta_{\mu\nu}\pi\,, \nn \\
l_{\mu\nu} &\to l_{\mu\nu} -  M_{\text{Pl}} \phi_0^2 \left( \beta_1^{\text{}} + 2  \beta_2^{\text{}} + \beta_3^{\text{}}  \right)\eta_{\mu\nu}\pi\,,
\label{confTrans}
\end{align}
where $\eta$ denotes the flat Minkowski metric as usual. Finally we can canonically normalise $\pi$ by sending
\be
\pi \to \frac{\pi}{\sqrt{3} M_{\text{Pl}}^{\text{}} (\beta_1^{\text{}} + 2 \beta_2^{\text{}} + \beta_3^{\text{}}) \phi_0^2}\,,
\ee
which then results in the fully demixed kinetic quadratic action
\be 
{\cal S}^{\rm kin}_2 = \int d^4x \left[-\frac{1}{2}\pa_\mu \delta\phi\pa^\mu \delta\phi -\frac{1}{2}\pa_\mu \pi \pa^\mu \pi + {\cal L}_2^{\rm EH}(h) +{\cal L}_2^{\rm EH}(l)\right].
\ee 

Finally we look at the potential interactions at quadratic order. After the replacements for tadpole cancellation, demixing kinetic modes and normalising the fields, we find 
\begin{align}
{\cal S}^{\rm pot}_2 &= \int d^4x \bigg[ \phi_0^2 (\beta_1^{\text{}} + 2 \beta_2^{\text{}} + \beta_3^{\text{}}) \bigg(\tfrac{1}{4} h_{\mn} h^{\mn} -  \tfrac{1}{4} h^{\mu}{}_{\mu} h^{\nu}{}_{\nu}
+ \tfrac{1}{4} l_{\mn} l^{\mn} -  \tfrac{1}{4} l^{\mu}{}_{\mu} l^{\nu}{}_{\nu}  \nn \\
&-    \tfrac{1}{2} h^{\mn} l_{\mn} + \tfrac{1}{2} h^{\mu}{}_{\mu} l^{\nu}{}_{\nu}  -  \sqrt{3} h^{\mu}{}_{\mu} \pi + \sqrt{3} l^{\mu}{}_{\mu} \pi - 4 \pi^2\bigg) - W_2(\delta\phi)\bigg].
\end{align}
%where $W_2(\delta\phi)$ denotes the quadratic potential contribution for $\delta\phi$ coming from $W(\Phi)$. 
The mass matrix between the different modes remains mixed just as in the standard bigravity case, with all residual terms proportional to powers of $\phi_0^2$. 
%(at least as long as the $\Phi$ potential is expressible as a polynomial and has no tadpole term itself, i.e. $W_2(\delta\phi)$ does not include a tadpole term for $\delta\phi$). 
Note that $\delta\phi$ is completely decoupled, but $h,l,\pi$ are all mixed.

%\\

%\ni {\bf Cubic interactions}:
%\ni {\bf Higher order interactions}:

\subsection{Non-linear interactions and strong coupling}

%\subsubsection{Cubic interactions}
\ni {\bf Cubic interactions}:
We can now finally move on to higher order interactions, which in particular will set the strong coupling scales of the theory and describe its true ``interaction structure''. The same tensor-scalar interactions (and resulting pure scalar interactions via \eqref{confTrans}) are present as for the standard bigravity theory. However, in addition new tensor-scalar and scalar-scalar interactions are present in our theory as well. 

In order to disentangle these two types of interactions, and their different physical properties, we will use two types of scaling limits. We begin by taking the following scaling limit, which will eliminate all cubic interactions involving tensors
\begin{align}
{\rm SL}_1:  \quad M_{\rm Pl} \to \infty\,.
\label{DL1}
\end{align}
This limit isolates pure scalar-scalar interactions at cubic order, which are given by 
\begin{align}
{\cal S}_{\rm {DL_1}}^{\rm 3-scalar} &= \int d^4x \left[ -8 (\beta_1^{\text{}} + 2 \beta_2^{\text{}} + \beta_3^{\text{}}) \phi_0 \delta \phi \pi^2 + 2 \frac{\delta \phi}{\phi_0} \pi \Box \pi - 
\frac{\left(\delta \phi (\Box \pi)^2 -  \delta \phi \pi_{\mu\nu} \pi^{\mu\nu}\right)}{6 (\beta_1^{\text{}} + 2 \beta_2^{\text{}} + \beta_3^{\text{}}) \phi_0^3}\right]
 \nn \\
&\sim \int d^4x \left[ \phi_0 \delta \phi \pi^2 + \frac{\delta \phi}{\phi_0} \pi \Box \pi  + \frac{\delta \phi}{\phi_0^3} (\Box \pi)^2 +  \frac{\delta \phi}{\phi_0^3} \pi_{\mu\nu} \pi^{\mu\nu} \right],
\label{cubic-DL1}
\end{align}
where we have suppressed constant dimensionless ${\cal O}(1)$ factors in going to the second line. These pure scalar interactions immediately underline the need for $\delta\phi \ll \phi_0$ in order for our perturbative approach to be valid. Otherwise e.g. the second term above immediately becomes larger than the (quadratic order) kinetic term for $\pi$, invalidating a perturbative expansion like ours here, which implicitly assumes that higher orders are subsequently more suppressed than lower orders (otherwise we in general need to keep track of arbitrarily large orders and can never truncate). Also note the first term, which is simply a non-derivative potential-type term, has not disappeared here since $\phi_0$ cannot simply be taken to zero without invalidating the perturbative approach. Note that one can, however, take $\phi_0 \to 0$ if one is willing to scale (and in principle eliminate) $\delta\phi$ at the same time (see below). As long as $\delta\phi \ll \phi_0$ and, as inspection of the above action shows, also $|\delta\phi (\Box \pi)^2| \ll |\phi_0^3 \pi_\mu\pi^\mu|$, the cubic action is under control. This is effectively a restriction on the validity of our local $\Phi \to \phi_0 + \delta\phi$ expansion. Since $\phi_0$ can be chosen arbitrarily, we can always (at least for a `short time') satisfy these conditions.\footnote{When expanding around the value taken by $\Phi$ at a given time, instantaneously (i.e. at that given time) $\delta\phi = 0$ and the inequality is trivially satisfied for any non-zero $\Phi$. How long the expansion around $\phi_0$ remains valid will depend on the evolution of $\Phi$ and hence on the choice of potential and mass interactions and coupling constants in the action. However, for a smoothly and continuously evolving $\Phi$ and hence $\delta\phi$ the expansion will always remain valid for a finite and non-zero length of time.} Even though it may therefore be tempting to turn these conditions into a new additional ``cutoff'', one should refrain from doing so, since this is purely a result of the initial choice of $\phi_0$ and a choice that satisfies the above inequalities can always be made.\footnote{For example, consider a monotonically growing $\phi$. Once the perturbative description around a given initially chosen $\phi_0$ becomes strongly coupled (since $\delta\phi$ becomes larger and larger and eventually dominates over $\phi_0$), we simply choose to expand around a new more `recent' $\hat\phi_0$, where $\phi_0 = \Phi(t_1)$, $\hat\phi_0 = \Phi(t_2)$ and $t_2 > t_1$, and can recover a valid perturbative description in the process. In this sense we suggest thinking of $\phi_0$ as a `reference value' - while it does instantaneously satisfy the background equations of motion for $\Phi$, $\phi_0$ once chosen and as defined by us here, has no dynamics (i.e. it is not a dynamical background variable). It is a constant reference value useful to keep track of the dominant normalising effects for the fields, but all the dynamics for $\Phi$ resides in $\delta\phi$.}

Secondly we consider another scaling limit, which essentially recovers the standard decoupling limit of bigravity. Here we eliminate the new dynamical scalar $\delta\phi$ altogether and afterwards (the ordering is important) take a scaling limit resembling the bigravity $\Lambda_3$ decoupling limit, where $\phi_0$ plays the role of the bigravity mass parameter $m$. The limit we take is therefore
\begin{align}
{\rm SL}_2&:  &\delta\phi &\to 0\,,&M_{\rm Pl} &\to \infty\,, &\phi_0 &\to 0\,, &\tilde \Lambda_3^3 &\equiv M_{\rm Pl} \phi_0^2 \to \text{ fixed},
\label{DL2}
\end{align}
where $\delta\phi \to 0$ before the remaining scaling limit is taken, as discussed above. 
Focussing on the scalar interactions that arise from the scalar-tensor interactions via the demixing transformation \eqref{confTrans}, we find 
\begin{align}
{\cal S}^{\rm 3-scalar}_{\rm DL_2} &= \int d^4x \left[\frac{(\beta_1^{\text{}} + 3 \beta_2^{\text{}} + 2 \beta_3^{\text{}})}{3 \sqrt{3} \tilde\Lambda_3^3 (\beta_1^{\text{}} + 2 \beta_2^{\text{}} + \beta_3^{\text{}})^2}
\left[\pi (\Box\pi)^2 - \pi \pi_{\mu\nu}\pi^{\mu\nu}\right]
 + \frac{\pi_\mu \pi^\mu \Box\pi}{2\sqrt{3} \tilde \Lambda_3^3 ( \beta_1^{\text{}} + 2  \beta_2^{\text{}} + \beta_3^{\text{}})} \right]\nn \\
 &\sim  
 \frac{1}{\tilde\Lambda_3^3}
\int d^4x \left[\pi (\Box\pi)^2 - \pi \pi_{\mu\nu}\pi^{\mu\nu}
 + \pi_\mu \pi^\mu \Box\pi \right],
 \label{cubic-DL2}
\end{align}
where we have suppressed constant dimensionless ${\cal O}(1)$ factors in going to the second line. These interactions unsurprisingly are precisely analogous to those found for standard bigravity, with $m \to \phi_0$ and an associated strong coupling scale $\tilde \Lambda_3$. 

Having considered two particular scaling interactions, let us now pull everything together and look at the complete set of interactions at cubic order involving scalars only (post-demixing via \eqref{confTrans}). We find 
\begin{align}
{\cal S}^{\rm 3-scalar} = {\cal S}^{\rm 3-scalar}_{\rm DL_1} + {\cal S}^{\rm 3-scalar}_{\rm DL_2} + \frac{2(\beta_3-\beta_1)}{\sqrt{3}\tilde \Lambda_3^3}\int d^4x \left[\frac{4\phi_0^4 \pi^3}{3} + \frac{\phi_0^2 \pi \pi_\mu \pi^\mu}{(\beta_1^{\text{}} + 3 \beta_2^{\text{}} + 2 \beta_3^{\text{}})}\right],
\label{fullcubic}
\end{align}
where the additional terms that are suppressed in the two scaling limits considered above are pure $\pi$ interactions suppressed by scales larger than $\tilde \Lambda_3$, as they would exist in standard bigravity as well. In summary,  the final result for the cubic action shows that we have the same cubic interactions as for standard bigravity, with $m \to \phi_0$, supplemented by perturbative corrections suppressed by $1/\phi_0^p$, where $p$ is some power $p \leq 3$. When $\phi_0$ is chosen such that our perturbative expansion is valid\footnote{Recall that this requires $\delta\phi \ll \phi_0$ as well as a restriction on the relation between $\delta\phi$ and $\phi_0^3$, as discussed above.}, $\tilde \Lambda_3$ therefore is the strong coupling scale of the theory, as may have been guessed naively. We emphasize that, just as the discussion of the quadratic action, this hinges on enforcing the tadpole cancellation requirements \eqref{betasol}. Otherwise a whole new host of interactions at different scales would apparently be present. 
\\

%\subsubsection{Higher order interactions}
\ni {\bf Higher order interactions}:
The structure we observed for cubic interactions is generic. Consider as a second explicit example interactions at quartic order. Taking our first scaling limit ${\rm SL}_1$ we then find
\begin{align}
{\cal S}_{\rm {DL_1}}^{\rm 4-scalar} &= \int d^4x \left[-4 (\beta_1^{\text{}} + 2 \beta_2^{\text{}} + \beta_3^{\text{}}) \delta \phi^2 \pi^2 
+ \frac{\delta \phi^2}{\phi_0^2}\left( 
\pi \Box\pi +  \frac{\left(\pi_{\mu\nu} \pi^{\mu\nu}-(\Box\pi)^2\right)}{12 (\beta_1^{\text{}} + 2 \beta_2^{\text{}} + \beta_3^{\text{}}) \phi_0^2} \right)\right] \nn \\
&\sim \int d^4x \left[-\delta \phi^2 \pi^2 + \frac{\delta \phi^2}{\phi_0^2}\left( \pi\Box\pi - \frac{(\Box\pi)^2}{\phi_0^2} + \frac{\pi_{\mu\nu}\pi^{\mu\nu}}{\phi_0^2}\right)\right],
\label{quartic-DL1}
\end{align}
where as before we have suppressed constant dimensionless ${\cal O}(1)$ factors in going to the second line, we have a potential-term like contribution and extra derivative interactions suppressed by powers of $\phi_0$. Note that the same conditions as for the quadratic and cubic interactions still ensure that our perturbative expansion is valid at this order.

Moving on to look at ${\rm SL}_2$ for quartic interactions we have 
\begin{align}
{\cal S}^{\rm 4-scalar}_{\rm DL_2} \sim 
\frac{1}{\tilde\Lambda_3^6}\int d^4x &\left[\pi (\Box\pi)^3 + \pi_\mu \pi^\mu (\Box\pi)^2 +  \pi^\mu \pi_{\mu\nu} \pi^\nu \Box\pi  \right. \nn \\
&+ \left. \pi \pi^{\mu\nu} \pi_{\nu\rho} \pi^{\rho}_{\mu} -  \pi \Box\pi \pi_{\mu\nu}\pi^{\mu\nu} - \pi_\rho \pi^\rho \pi_{\mu\nu} \pi^{\mu\nu} \right]\,,
\label{quartic-DL2}
\end{align}
where we now suppress constant dimensionless ${\cal O}(1)$ factors from the start to avoid clutter and we can read off the effective strong coupling scale $\tilde \Lambda_3$ again and see the same type of interactions as for the standard bigravity decoupling limit. Note that the pure-scalar interactions from before de-mixing, which come in at lower scales ($\tilde \Lambda_5$ for cubic order, $\tilde \Lambda_4$ for quartic order etc.) cancel up to total derivatives due to the anti-symmetric structure of the interaction potential, just as for standard massive and bi-gravity. 

The above interactions are supplemented by other scalar interactions that vanish in the limits considered above, i.e. suppressed by scales larger than $\tilde \Lambda_3$ and/or powers of $\phi_0$. All the interactions are (Boulware-Deser) ghost-free, as shown by the constraint analyses \cite{Huang:2012pe}, even though the corresponding equations of motion at higher orders naively (i.e. without applying additional transformations) become higher-order in derivatives (and hence naively lead to Ostrogradski instabilities), just as for the standard bi- and multi-gravity cases \cite{deRham:2013hsa,Noller:2015eda}. 

Given that this overall structure stays in place also at other generic higher orders, it makes sense to write
\be
\Lambda_{\text{strong coupling}} = \tilde \Lambda_3,
\ee
which is the scale where perturbative unitarity is lost\footnote{Note that, while the Vainshtein scale also descends from this, there will be an extra dependence e.g. on the mass of the object around which we investigate screening, just as in massive gravity \cite{Hinterbichler:2011tt}.}. We re-emphasise that this result is highly non-trivial, given that the terms with the new dynamical scalar $\Phi = \phi_0 + \delta\phi$ change the structure of interactions (by providing additional vertices) and could have changed the relevant suppressing energy scales as well. What was crucial for $\tilde \Lambda_3$ to become the effective strong coupling scale was that the extra scalar $\delta\phi$ decoupled at quadratic order due to the tadpole cancellation conditions \eqref{betasol}, so that no extra de-mixing at this order was necessary and the field normalisations therefore stayed as they were in standard bigravity.

\subsection{Vainshtein screening and the equations of motion}

In order to see what phenomenological effects the new scaling limit interactions might have, we now compute the contribution to the $\pi$ equations of motion from the different limits presented above. We do so in the case of a spherically symmetric and static field configuration (e.g. around a central point-like matter source), in close analogy to what is done for galileons \cite{Nicolis:2008in}. We focus on the cubic order contributions and find that the contribution to the $\pi$ equations of motion coming from \eqref{cubic-DL1} is
\begin{align}
{\cal E}^{\rm 3-scalar}_{\rm DL_1} &= \phi_0 \delta\phi \pi + \frac{\delta\phi}{\phi_0} \pi^{'} + \frac{\delta\phi^{'}}{\phi_0} \pi^{'} + \frac{\delta\phi^{'}}{\phi_0^3 r^2} \pi^{'}  + \frac{\delta\phi^{''}}{\phi_0^3 r} \pi^{'}  + \frac{\delta\phi}{\phi_0} \pi^{''} \nn \\ &+ \frac{\delta\phi^{'}}{\phi_0^3 r}\pi^{''}+  \frac{\delta\phi^{''}}{\phi_0^3}\pi^{''} + \frac{\delta\phi}{\phi_0^3 r}\pi^{'''} + \frac{\delta\phi^{'}}{\phi_0^3}\pi^{'''} + \frac{\delta\phi}{\phi_0^3}\pi^{''''},
\label{Eom-cubic-DL1}
\end{align}
where we have suppressed constant ${\cal O}(1)$ dimensionless constants from the start this time and ${}^{'} \equiv \partial/\partial_r$. From \eqref{cubic-DL2} we have 
\begin{align}
{\cal E}^{\rm 3-scalar}_{\rm DL_2} &= \frac{1}{\tilde\Lambda_3^3} \left({\pi^{''}}^2  + \pi^{'}\pi^{'''} + \pi\pi^{''''}
+ \frac{\pi \pi^{'''}}{r}
+ \frac{\pi^{'}\pi^{''}}{r}
+ \frac{{\pi^{'}}^2}{r^2} 
\right).
\label{Eom-cubic-DL2}
\end{align}
The contribution seen in \eqref{Eom-cubic-DL2} are exactly as for standard bigravity with $m \to \phi_0$, as expected. Note the higher-derivative nature of the equations of motion -- this does not lead to an Ostrogradski ghost, since the cubic order action is complemented by infinitely many higher order terms and the full action written in this way is therefore degenerate. We will discuss this issue at the level of the action in the following subsection. The new terms in \eqref{Eom-cubic-DL1}, suppressed by powers of $\phi_0$ as expected, give new non-linear contributions in the spherically symmetric and static case considered here. This is consistent with the standard Vainshtein screening (since we can make $\delta\phi$ arbitrarily small for the initial evolution from any point onwards by choosing $\phi_0$ appropriately), but the terms in \eqref{Eom-cubic-DL1} will modify the non-linear background solution for $\pi$ and hence also modify the Vainshtein radius and screening effects. How this takes place again will be highly dependant on the evolution of $\Phi$ and hence on the choice of potential and initial conditions for $\Phi$.

\subsection{The massive gravity limit}

Having discussed the interaction structure for our ``scale free'' model of bigravity above, we can easily deduce what the interaction structure would be for an analogous model of scale-free massive gravity, which would be a particular model of the so-called ``mass-varying massive gravity'' type \cite{Huang:2012pe}. It corresponds to freezing one of the dynamical vielbeins in \eqref{vielbeinAction} - for definiteness (this is an arbitrary choice) we freeze $E_{(2)}$ by sending $E_{(2)}{}^{A}_{\mu} \to \delta^A_\mu$ or equivalently by setting $g^{(2)}_{\mu\nu} = \eta_{\mu\nu}$ in \eqref{metricAction}. We still introduce \St{} fields via this now fixed reference metric. 
We emphasise that what we mean by the ``massive gravity limit'' here literally consists of freezing one metric/vielbein and we do not try to obtain this limit as a decoupling limit of the full action, but we do keep the $\beta_4$ term, which now becomes a simple standard mass term for $\Phi$ in flat space.
In the following we briefly go through the same steps as above to show what the interaction structure of the corresponding scale-free massive gravity model is and what changes in comparison to the bigravity case considered above.
\\

\ni {\bf Tadpoles and the quadratic action}: 
%\subsubsection{Tadpole cancellations}
Inspection of \eqref{tadpoles} suggests that the tadpole conditions do not change in the (single) massive gravity case and an explicit check verifies that we still require 
\begin{align}
\beta_0 &= -3\beta_1 - 3\beta_2 - \beta_3\,,\\
\beta_4 &= -\beta_1 - 3\beta_2 - 3\beta_3\,,
\end{align}
for the linear tadpole terms to cancel. Moving on to the quadratic action, we see that scalar-scalar mixing is again forbidden by implementing the tadpole conditions (otherwise it would still take on the form \eqref{betasol}). Scalar-tensor mixing at quadratic order is still eliminated by the linearised conformal transformation
\begin{align}
h_{\mu\nu} &\to h_{\mu\nu} +  M_{\text{Pl}} \phi_0^2 \left( \beta_1^{\text{}} + 2  \beta_2^{\text{}} + \beta_3^{\text{}}  \right)\eta_{\mu\nu}\pi\,,
\end{align}
and of course no transformation for the second tensor $(l)$ is needed any more, since $l$ is not a dynamical degree of freedom in massive gravity. Finally we can canonically normalise $\pi$ by sending
\be
\pi \to \frac{\pi}{\sqrt{3/2} M_{\text{Pl}}^{\text{}} (\beta_1^{\text{}} + 2 \beta_2^{\text{}} + \beta_3^{\text{}}) \phi_0^2}\,,
\ee
which then results in the fully demixed kinetic quadratic action and where we note that we have an extra factor of $\sqrt{2}$ in comparison to the bigravity case, owing to the fact that we only demixed from one and not from two tensors.
\\

%\subsubsection{Cubic interactions}
\ni {\bf Non-linear interactions (cubic)}:
We will again utilise the two scaling limits ${\rm SL}_{1}, {\rm SL}_{2}$ defined above in order to disentangle interactions. At cubic order we now have
\begin{align}
{\cal S}_{\rm {DL_1}}^{\rm 3-scalar} &= \int d^4x \left[ -4 (\beta_1^{\text{}} + 2 \beta_2^{\text{}} + \beta_3^{\text{}}) \phi_0 \delta \phi \pi^2 + 2 \frac{\delta \phi}{\phi_0} \pi \Box \pi - 
\frac{\left(\delta \phi (\Box \pi)^2 -  \delta \phi \pi_{\mu\nu} \pi^{\mu\nu}\right)}{3 (\beta_1^{\text{}} + 2 \beta_2^{\text{}} + \beta_3^{\text{}}) \phi_0^3}\right],
\end{align}
where nothing has changed in comparison with the bigravity case except for some numerical factors due to the changed normalisation of $\pi$ for the quadratic action. However, for our second scaling limit, i.e. the one resembling the standard bigravity decoupling limit, we have
\begin{align}
{\cal S}^{\rm 3-scalar}_{\rm DL_2} &= \int d^4x \frac{( \beta_2^{\text{}} + \beta_3^{\text{}})}{3 \sqrt{3/2} \tilde\Lambda_3^3 (\beta_1^{\text{}} + 2 \beta_2^{\text{}} + \beta_3^{\text{}})^2}
\left[\pi (\Box\pi)^2 - \pi \pi_{\mu\nu}\pi^{\mu\nu}\right].
\end{align}
We notice two differences when comparing with \eqref{cubic-DL2}: (1) Firstly the absence of a term like $\pi_\mu\pi^\mu\Box\pi$. This is due to the fact that now we are \Sting{} a fixed reference metric, whereas previously we had to Taylor-expand $E_{(2)}$ post-\Sting{}, which gave rise to a non-local dependence on $\pi$ via terms such as the one missing here.\footnote{By `non-local dependence' we mean that this introduced infinitely many interaction terms for $\pi$ at arbitrarily high orders in derivatives and fields. The $\pi_\mu\pi^\mu\Box\pi$ term discussed here is the cubic order term from that infinite expansion.} Note that at cubic order scalar-tensor mixing, and hence the appearance of terms like the one missing here, could also be removed by a local field re-definition in the bigravity case, but at higher orders this is not the case. We will see the difference between the scale-free massive and bi-gravity cases related to these terms even more clearly at quartic order below. (2) The now non-dynamical nature of $E_{(2)}$ also leads to a different $\beta$-dependence when compared with \eqref{cubic-DL2}. Unsurprisingly the interactions found in the ${\rm SL}_{2}$ limit here are precisely analogous to those found for standard massive gravity, with $m \to \phi_0$ and an associated strong coupling scale $\tilde \Lambda_3$. 
Pulling everything together at cubic order we find that the complete set of interactions at this order is 
\begin{align}
{\cal S}^{\rm 3-scalar} &= {\cal S}^{\rm 3-scalar}_{\rm DL_1} + {\cal S}^{\rm 3-scalar}_{\rm DL_2}- \nn \\ &- \frac{2}{\sqrt{3/2}\tilde \Lambda_3^3}\int d^4x \left[\frac{2(2\beta_1 + 3\beta_2+\beta_3)\phi_0^4 \pi^3}{3} + \frac{(\beta_1+\beta_2)\phi_0^2 \pi \pi_\mu \pi^\mu}{(\beta_1^{\text{}} + 2 \beta_2^{\text{}} + \beta_3^{\text{}})}\right],
\end{align}
where comparison with \eqref{fullcubic} reveals a modified $\beta$-dependence compared with the bigravity case as discussed above. 
\\

\ni {\bf Non-linear interactions (quartic)}:
Moving on to quartic interactions, in our first scaling limit ${\rm SL}_1$ we find
\begin{align}
{\cal S}_{\rm {DL_1}}^{\rm 4-scalar} &= \int d^4x \left[-2 (\beta_1^{\text{}} + 2 \beta_2^{\text{}} + \beta_3^{\text{}}) \delta \phi^2 \pi^2 
+ \frac{\delta \phi^2}{\phi_0^2}\left( 
\pi \Box\pi +  \frac{\left(\pi_{\mu\nu} \pi^{\mu\nu}-(\Box\pi)^2\right)}{6 (\beta_1^{\text{}} + 2 \beta_2^{\text{}} + \beta_3^{\text{}}) \phi_0^4} \right)\right] \,.
\end{align}
In this limit we again see no differences to the bigravity case except for numerical factors coming from the slightly different normalisation for $\pi$. Differences to the bigravity case are more pronounced in the ${\rm SL}_2$ scaling limit, where we obtain
\begin{align}
{\cal S}^{\rm 4-scalar}_{\rm DL_2} = \frac{\beta_3}{27 \tilde \Lambda_3^6 (\beta_1^{\text{}} + 2 \beta_2^{\text{}} + \beta_3^{\text{}})^3} \int d^4x \left[
\pi(\Box\pi)^3 + 2 \pi \pi^{\mu\nu} \pi_{\nu\rho} \pi^{\rho}_\mu - 3\pi \Box\pi \pi_{\mu\nu} \pi^{\mu\nu}\right]\,,
\end{align}
and comparison with \eqref{quartic-DL2} empasises the point discussed for cubic interactions above. Namely that additional higher-derivative interactions coming from the dynamical nature of both vielbeins in the bigravity case are absent in the massive gravity case. As before, at quartic order the above interactions are supplemented by other scalar interactions that vanish in the limits considered above, i.e. suppressed by scales larger than $\tilde \Lambda_3$ and/or by powers of $\phi_0$.

\section{Quadratic action on generic backgrounds}\label{sec-Quad}

In the previous section we have investigated the interaction structure of scale-free bigravity around flat-space configurations for the metrics and a constant configuration for the scalar field. We will now turn to analyze the structure of the quadratic action of the model, expanded around generic background configurations. To derive the quadratic action for generic backgrounds, we generalize the method introduced in \cite{Guarato:2013gba} for massive gravity and applied in \cite{Cusin:2015tmf} to the  (standard) bigravity case. We then specialize to homogeneous and isotropic backgrounds (FLRW) for the metrics and we write down the most general parametrization for the quadratic lagrangian in this context. We conclude  commenting on the background evolution of the model in the cosmological ansatz. 

The perturbed metrics are defined as
\be
g_{\mu\nu} = \bar g_{\mu\nu}+\mathbb{h}_{\mu\nu}\equiv \bar g_{\mu\nu}+\frac{h_{\mu\nu}}{M_{\text{Pl}}}\,, 
\ee
\be
f_{\mu\nu} = \bar f_{\mu\nu}+\mathbb{l}_{\mu\nu}\equiv \bar f_{\mu\nu}+\frac{\ell_{\mu\nu}}{M_{\text{Pl}}}\,, 
\ee
where $\bar g_{\mu\nu}$ and $\bar f_{\mu\nu}$ indicate generic background solutions and $h_{\mu\nu}$ and $\ell_{\mu\nu}$ are canonically normalized variables. 
From now on,  the indices of the tensor $\mathbb{h}_{\mu\nu}$  and $h_{\mu\nu}$ will be raised and lowered with the physical background metric $\bar g_{\mu\nu}$, whereas the indices of the tensor $\mathbb{l}_{\mu\nu}$  and $\ell_{\mu\nu}$will be raised and lowered with the background metric $\bar f_{\mu\nu}$. The scalar field is expanded around a background configuration as
\be
\Phi=\phi+\delta \phi\,.
\ee
We underline that in the equation above, $\phi$ is a dynamical field,  solution of the background equation of motion for the scalar field $\Phi$.\footnote{We emphasize that our approach here is different from that of section \ref{sec-int}. There, in the context of studying the interaction structure of the theory around Minkowski backgrounds,  we considered the split $\Phi=\phi_0+\delta\phi$, with $\phi_0$ being a \emph{fixed} reference configuration for the scalar field.}

Using the method illustrated in appendix \ref{perturbed mass} and based on the results of \cite{Cusin:2015tmf}, we derive the general expression for the perturbed action, quadratic in the canonically normalized metric perturbations  $h_{\mu\nu}$ and $\ell_{\mu\nu}$ and $\delta \phi$
\be\label{eqcovv}
\mathcal{S}_{2} =\mathcal{S}^{kin}_{2} +\mathcal{S}^{m}_{2} \,,
\ee
\small
\be\label{eqkin}
\mathcal{S}_{2}^{kin} =\frac{1}{2}\int d^4x \,\sqrt{- \bar g}\, \left(h_{\mu\nu} \mathcal{E}^{\mu\nu\alpha\beta}(\bar{g})h_{\alpha\beta}-(\nabla_{\mu} \delta\phi)^2\right)+\frac{1}{2}\int d^4x\,\,\sqrt{- \bar f}\,\, \lc_{\mu\nu} \mathcal{E}^{\mu\nu\alpha\beta}(\bar{f})\lc_{\alpha\beta}\,, 
\ee
\begin{align}\label{Sm}
\mathcal{S}_{2}^{m}= -&\frac{1}{2}\int d^4x\,\sqrt{- \bar g}\,\phi^2\,\Big[\MM^{\mu\nu\alpha\beta}_{\gc\gc}(\bar f,\bar g)h_{\mu\nu}h_{\alpha\beta} + \MM^{\mu\nu\alpha\beta}_{\gc\fc}(\bar f,\bar g)h_{\mu\nu}\lc_{\alpha\beta} +\MM^{\mu\nu\alpha\beta}_{\fc\fc}(\bar f,\bar g)\lc_{\mu\nu}\lc_{\alpha\beta}\Big] -\nn\\
-&\frac{ M^2_{\text{Pl}}}{2}\int d^4x\,\sqrt{-\bar g}\,\phi\,\Big[2 \MM^{\mu\nu}_{\gc}(\bar f, \bar g)h_{\mu\nu}\,\delta\phi +2 \MM^{\mu\nu}_{\fc}(\bar f, \bar g)\lc_{\mu\nu}\,\delta\phi\Big] -\,\nn\\
-&\frac{M^2_g}{2}\int d^4x\,\sqrt{- \bar g}\, \,V(\bar{g},\bar{f})\,\delta\phi^2 -\frac{1}{2}\int d^4x\,\sqrt{-\bar g}\, \,\frac{\partial^{(2)} W}{\partial \Phi\partial\Phi}\bigg|_{g=\bar{g}\,,f=\bar{f}}\,\,\delta\phi^2\,,
\end{align}\normalsize
where the tensors $\mathcal{M}^{\mu\nu\alpha\beta}_{\bullet \bullet}$ and $\mathcal{M}_{\bullet \bullet}^{\mu\nu}$ are defined in appendix \ref{perturbed mass} and $\EE^{\mu\nu\al\beta}(\bar g)$ is the Lichnerowicz operator in curved space-time, whose explicit expression can be found in appendix \ref{perturbed mass}. 

We observe that when the massive gravity limit $\ell_{\mu\nu}\rightarrow 0$ is taken, we recover a specific implementation of the mass-varying massive gravity model proposed  in  \cite{Huang:2012pe}. The resulting mass term in this limit is the one derived in \cite{Guarato:2013gba} for (standard) massive gravity, with an additional contribution mixing the scalar field with the metric perturbation $h_{\mu\nu}$.

\subsection{Cosmological background case}\label{sec-FRW}

We now want to specialize our results for the quadratic action (\ref{eqcovv}) to the case of cosmological backgrounds. The kinetic structure of the linearized theory is standard (two copies of GR plus a scalar field). We will focus on the parametrization of the mass term in the case of homogeneous and isotropic background solutions. 
\\

%\subsubsection{Cosmological ansatz and  background equations}\label{cosmology}

\ni{\bf Cosmological ansatz}:
We consider solutions of bigravity where both metrics are spatially isotropic and homogeneous. For simplicity, we also assume that both metrics have  flat spatial sections, $K=0$. Modulo time re-parameterizations,  the most general form for the metrics (in conformal time $\tau$) is
\be\label{FRWg}
\bar{g}_{\mu\nu}dx^{\mu}dx^{\nu}=a^2(\tau)\left(-d\tau^2+\delta_{ij}dx^idx^j\right)\,,
\ee
\be\label{FRWf}
\bar{f}_{\mu\nu}dx^{\mu}dx^{\nu}=b^2(\tau)\left(-c^2(\tau) d\tau^2+\delta_{ij}dx^idx^j\right)\,.
\ee
Here $a$ and $b$ are the scale factors of the two metrics and $c$ is a lapse function for $f$.
It is convenient to define both the conformal Hubble parameter ($\HH$) and the standard one (${H}$) for both metrics
\be\label{Hubble-definition}
H=\frac{\mathcal{H}}{a}=\frac{a'}{a^2}\,,\hspace{0.5 cm} H_f=\frac{\mathcal{H}_f}{b}=\frac{b'}{b^2\,c}\,,
\ee
where $'$  denotes the derivative with respect to the conformal time $\tau$. We introduce also the ratio between the two scale factors
\be
r=\frac{b}{a}\,.
\ee
We indicate with $\phi$ the background value of the scalar field $\Phi$. 

In the matter sector, we consider the energy-momentum tensor of a covariantly conserved perfect fluid with equation of state $p=w \rho$ and 4-velocity $u^{\mu}$. Explicitly,
\begin{align}
&T_{\mu\nu}^{(m)}=\left(p+\rho\right)\, u_{\mu} u_{\nu}+p \,g_{\mu\nu}\,,\\
&\rho'=-3(\rho+p)\,\mathcal{H}\,,\\
&p=w\rho\,.
\end{align}
\\

\ni{\bf Background equations}:
The equation of motion for the background value of the scalar field, $\phi\equiv \phi(\tau)$ can be written as
\be\label{eqinf}
\phi''+2\mathcal{H}\phi'+a^2\,W_{,\phi}+a^2\,M_{\text{Pl}}^2\,V(r,c)\, \phi=0\,,
\ee
where
\be\label{mu}
V(r,c)\equiv \bar{V}(\bar{g},\bar{f})= \beta_0+\beta_1\left(c+3\right) r+3\, \beta_2 (c+1) r^2+\beta_3 \left(1+3c\right)r^3+\beta_4\, c \,r^4\,.
\ee
It is useful to introduce an \emph{effective potential}
\be\label{effpotential0}
\mathcal{W}(\phi, r, c)\equiv W(\phi)+\frac{1}{2}M_{\text{Pl}}^2\,V(r,c)\,\phi^2\,.
\ee
The quantity $V(r,c)$ gives a time-dependent correction to the scalar field mass. We observe that if we set the potential of the scalar field to zero from the very beginning, $W=0$,  we still get a quadratic potential from the coupling to the matter sector, $\mathcal{W}=1/2 M_{\text{Pl}}^2\,V\phi^2$.\footnote{We observe that our model can be easily generalized promoting in eq. (\ref{metricAction}) the factor appearing in front of the bigravity potential to a generic function of the scalar field $\Phi$, i.e. $\Phi^2\rightarrow P(\Phi)$. In this case setting $W=0$ we generate an effective potential for the scalar field given by the function $P(\Phi)$}

As already shown, the Bianchi constraint (\ref{b1}) is equivalent to the covariant conservation of the total energy momentum tensor. In the cosmological ansatz it can be written as 
\be\label{e:bian}
\rho_{tot}'=-3\mathcal{H}\,\left(\rho_{tot}+p_{tot}\right)\,,
\ee
where $\rho_{tot}=\rho_{m}+\rho_{\phi}$ and $\rho_{\phi}$ is the energy density of the scalar field. Explicitly
\be\label{e:rhophi}
\rho_{\phi}=-\left(T_{\Phi}\right)^{0}_0=%-\left(\mathcal{T}_{\Phi}\right)^{0}_0+M_{\text{Pl}}^2\,\Phi^2\,\mathcal{M}^0_0(g)=
\frac{\phi'^2}{2 a^2}+W(\phi)+\frac{\phi^2}{16\pi G}\left(\beta_3\, r^3+3\beta_2\,r^2+3\beta_1\,r+\beta_0\right) \,.
\ee
The associated pressure $p_{\phi}\equiv\omega_{\phi}\,\rho_{\phi}$ is given by 
\be\label{state}
p_{\phi}=\frac{1}{3}\left(T_{\Phi}\right)^{i}_i=%\frac{1}{3}\,\left(\mathcal{T}_{\Phi}\right)^{i}_i-\frac{1}{3}\,M_{\text{Pl}}^2\,\Phi^2\,\mathcal{M}^i_i(g)=
\frac{\phi'^2}{2 a^2}-W(\phi)-\frac{\phi^2}{16\pi G}\left(\beta_3 c \,r^3+\beta_2(2c+1)r^2+\beta_1(c+2)r+\beta_0\right) \,.
\ee
Note that both $p_\phi$ and $\rho_\phi$ include contributions coming from the bigravity potential, which would still be relevant if $\phi$ were not a dynamical field, but just a fixed mass scale. In the limit when $\phi$'s evolution is frozen this will be the dominant contribution together with the stationary value of $W(\phi)$, so in a slight abuse of notation we will still refer to these contributions via $p_\phi$ and $\rho_\phi$ even when there is (effectively) no dynamical $\phi$.
It is easy to show that the Bianchi constraint eq. (\ref{b2}) is equivalent to
\be\label{BC}
2\phi'\left(\beta_1+3\beta_2r+3\beta_3 r^2+\beta_4 r^3\right)+3\phi\left(\mathcal{H}-\mathcal{H}_f\right)\left(\beta_1+2\beta_2r+\beta_3 r^2\right)=0\,,
\ee
which reduces to the standard constraint in bi-gravity for constant $\phi$ e.g. see equation (59) in \cite{Cusin:2015tmf}. We distinguish in the following two branches of solutions according to how the Bianchi constraint (\ref{BC}) is realized. We can either implement the constraint extracting e.g. the lapse $c$ or asking that the combinations of $\beta_i$ and $r$ in the round parenthesis are vanishing. Explicitly
\vspace{1 em}
\begin{align}\label{analytic}
&\text{\emph{ First Branch}}:\hspace{3 em}\left(\beta_1+3\beta_2r+3\beta_3 r^2+\beta_4 r^3\right)=0\,,\hspace{1 em}\left(\beta_1+2\beta_2r+\beta_3 r^2\right)=0\,,\\
&\nn\\
\vspace{2 em}
&\text{\emph{Second Branch}}:\hspace{3 em}(1-c)\,\mathcal{H}_f=-\frac{r'}{r}+\frac{2\, \phi'}{3\, \phi}\, \frac{\beta_1+3\beta_2r+3\beta_3 r^2+\beta_4 r^3}{\beta_1+2\beta_2r+\beta_3 r^2}\,.\label{lapse}
\end{align}
The first branch is the analogue of the algebraic branch in the standard bi-gravity formulation while the second one corresponds to the so called dynamical branch. For standard bi-gravity, the existence of two branches of solutions has been pointed out for the first time in \cite{Comelli:2011zm}. In the standard case, the evolution of perturbations in the second branch has been intensively studied in \cite{Cusin:2014psa, Cusin:2015pya, Lagos:2014lca, Konnig:2013gxa, Konnig:2014xva} while the evolution of tensor perturbations in the first branch is presented in \cite{Cusin:2015tmf}. In the next section we will analyze the main features of the two branches (\ref{analytic}) and (\ref{lapse}). 
 
The equations of motion (the Friedman equation and the acceleration equation) for the metric $g$  are given by
\be\label{F11}
3H^2=8\pi G\,\left(\rho+\rho_{\phi}\right)\,,
\ee
\be\label{F12}
3H^2+\frac{2H'}{a}=-8\pi G\,\left(p+p_{\phi}\right)\,,
\ee
while for the $f$ metric we find the equations of motion
\be\label{F21}
3\,H_f^2=\frac{\phi^2}{2}\left(\frac{\beta_1}{r^3}+\frac{3\beta_2}{r^2}+\frac{3\beta_3}{r}+\beta_4\right)\,,
\ee
\be\label{F22}
\frac{2\,H_f'}{b}=\frac{\phi^2}{2}\cdot\frac{(1-c)}{r^3}\cdot\left( \beta_1+2\beta_2r+\beta_3 r^2\right)\,.
\ee

%\be\label{F23}
%3\,H_f^2+\frac{2\,H_f'}{a\,c\, r}=\frac{\phi^2}{2}\left(\frac{1}{c\,r^3} (\beta_1+2\beta_2r+\beta_3 r^2)+\frac{1}{r^2} (\beta_2+2\beta_3 r+\beta_4 r^2)\right)\,.
%\ee
%\\

%where with a simple algebraic calculation one has 

%\textcolor{red}{GC: Do we really need the last expression above?}

%\subsubsection{First branch}

\ni{\bf First branch}:
 In the first branch the ratio between the two scale factors is constant $r=\bar{r}$ and there are the following two constraints
\be\label{firstbranch}
\left(\beta_1+3\beta_2\bar{r}+3\beta_3 \bar{r}^2+\beta_4 \bar{r}^3\right)=0\,,\hspace{2 em}\left(\beta_1+2\beta_2\bar{r}+\beta_3 \bar{r}^2\right)=0\,.
\ee
We assume one can solve both the equations simultaneously. This imposes a relation between/restriction on the $\beta_i$ and $\hat r$, which would not be satisfied by an arbitrary choice of $\beta_i$.  However, by looking at (\ref{F21},\ref{F22}) we see that the above constraints impose $H_f=0$  and $H'_f=0$ respectively. 
Using the definition (\ref{Hubble-definition}) this results in a constant $b$. Consequently for $r=\bar{r}=cnst$ we have a constant scale factor $a$ and thus $H=0$. This branch is therefore not viable to describe (homogeneous and isotropic) background cosmology. 
\\

%Note that this conclusion is based on assuming both metrics are spatially isotropic and homogeneous. Obviously by relaxing these assumptions there is a chance for a modified algebraic branch and viable cosmological solutions.\textcolor{red}{ GC: Isn't this obvious?}

%\subsubsection{Second branch}

\ni{\bf Second branch}:
In this branch (\ref{lapse}), the Bianchi constraint (\ref{BC}) can be used to extract the lapse $c$. We get 
\be\label{c}
c=\frac{1}{r}\cdot\frac{3\phi\left(r'+r\mathcal{H}\right)\left(\beta_1+2\beta_2r+\beta_3 r^2\right)}{2\phi'\left(\beta_1+3\beta_2r+3\beta_3 r^2+\beta_4 r^3\right)+3\phi\mathcal{H}\left(\beta_1+2\beta_2r+\beta_3 r^2\right)}\,.
\ee
It follows that in general in this branch the correction to the mass of the scalar field is a time-dependent quantity $V(r, c)$ given by eq. (\ref{mu}).

The Friedmann equations in this branch are given by eqs. (\ref{F11}) and (\ref{F21}) with 
\begin{align}
\rho_{\phi}&=\frac{\phi'^2}{2 a^2}+W(\phi)+\frac{\phi^2}{16\pi G}\left(\beta_3\, r^3+3\beta_2\,r^2+3\beta_1\,r+\beta_0\right) \,,\label{state2}\\
p_{\phi}&=\frac{\phi'^2}{2 a^2}-W(\phi)-\frac{\phi^2}{16\pi G}\left(\beta_3 c \,r^3+\beta_2(2c+1)r^2+\beta_1(c+2)r+\beta_0\right) \label{state22}\,.
\end{align}
The state parameter $\omega_{\phi}\equiv p_{\phi}/\rho_{\phi}$ is given by
\be\label{eq-state}
\omega_{\phi}=-1+\frac{\frac{\phi'^2}{a^2}+(1-c)\,r\,\frac{\phi^2}{16\pi G}(\beta_3\, r^2+2\beta_2\,r+\beta_1)}{\frac{\phi'^2}{2\,a^2}+W(\phi)+\frac{\phi^2}{16\pi G}(\beta_3\, r^3+3\beta_2\,r^2+3\beta_1\,r+\beta_0)}\,.
\ee
If the scalar field $\phi$ is slowly varying, the background evolution in this second branch will be very close to the one of standard bigravity in the dynamical branch.

\subsection{Mass term on cosmological backgrounds}\label{massesF}

With the ansatz (\ref{FRWg}, \ref{FRWf}) for the background metrics, homogeneity and isotropy request that the tensors $\MM_{\bullet\bullet}^{\al\beta\mu\nu}$ and $\MM_{\bullet}^{\mu\nu}$ in eq. (\ref{Sm}) admit the following general parametrization. For the $\gc\gc$ and $\fc\fc$ terms of $\MM_{\bullet\bullet}^{\alpha\beta\mu\nu}$ we have
\begin{align}
\MM^{0000}_{\bullet \bullet}(\bar f,\bar g)&=a^{-4} \alpha_{\bullet}(\tau)\,,\\
\MM^{ij00}_{\bullet \bullet}(\bar f,\bar g)&= \MM^{00ij}_{\bullet \bullet}(\bar f,\bar g)= a^{-4}  \gamma_{\bullet}(\tau) \delta^{ij}\,,\\
\MM^{i0j0}_{\bullet \bullet}(\bar f,\bar g)&=\MM^{0i0j}_{\bullet \bullet}(\bar f,\bar g)=a^{-4} \epsilon_{\bullet}(\tau) \delta^{ij}\,,\\
\MM^{ijkl}_{\bullet \bullet}(\bar f,\bar g)&=a^{-4}\left\{\eta_{\bullet}(\tau)\delta^{ij}\delta^{kl}+\frac{\sigma_{\bullet}(\tau)}{2}\left(\delta^{ik}\delta^{jl}+\delta^{il}\delta^{jk}\right)\right\}\,,
\end{align}
where $\bullet$ stands for either $\gc$ or $\fc$. For the mixed terms $\gc\fc$,  the parametrization takes the form
\begin{align}
\MM^{0000}_{\gc\fc}(\bar f,\bar g)&=a^{-4} \alpha_{\gc\fc}(\tau)\,,\\
\MM^{ij00}_{\gc\fc}(\bar f,\bar g)&=a^{-4}  \gamma_{\gc\fc}(\tau) \delta^{ij}\,,\\
\MM^{00ij}_{\gc\fc}(\bar f,\bar g)&=a^{-4} \gamma_{\fc\gc}(\tau) \delta^{ij} \,,\\
\MM^{i0j0}_{\gc\fc}(\bar f,\bar g)&=\MM^{0i0j}_{\gc\fc}(\bar f,\bar g)=a^{-4} \epsilon_{\gc\fc}(\tau) \delta^{ij}\,,\\
%\MM^{j0i0}_{hl}(\bar f,\bar g)&=-m^2 \delta^{ij}\epsilon_{lh}(\tau)\,,\\
\MM^{ijkl}_{\gc\fc}(\bar f,\bar g)&=a^{-4}\left\{\eta_{\gc\fc}(\tau)\delta^{ij}\delta^{kl}+\frac{\sigma_{\gc\fc}(\tau)}{2}\left(\delta^{ik}\delta^{jl}+\delta^{il}\delta^{jk}\right)\right\}\,.
\end{align}
For the tensors $\MM_{\bullet}^{\mu\nu}$ the parametization takes the form
\begin{align}
\MM_{\bullet}^{00}(\bar f,\bar g)&=a^{-2}\zeta_{\bullet}(\tau)/2\,,\\
\MM_{\bullet}^{ij}(\bar f,\bar g)&=a^{-2}\xi_{\bullet}(\tau)/2\,\,\,\delta^{ij}\,,
\end{align}
where $\bullet$ stands for either $\gc$ or $\fc$.

The functions $\alpha_{\bullet}$, $\gamma_{\bullet}$, $\epsilon_{\bullet}$, $\sigma_{\bullet}$, $\eta_{\bullet}$ (with $\bullet=\gc, \fc, \gc\fc$ or $\fc\gc$) and $\zeta_{\bullet}$, $\xi_{\bullet}$ (with $\bullet=\gc, \fc$) depend on conformal time through the ratio between the two scale factors, $r$, and the lapse function $c$.  Their explicit expressions are  given in appendix~\ref{FRW}. Note that contrary to $\gc\gc$ and $\fc\fc$, $\MM^{ij00}_{\gc\fc}\neq \MM^{00ij}_{\gc\fc}$ and we have introduced  $\gamma_{\gc\fc}\neq\gamma_{\fc\gc}$.

Given this parametrization it is straightforward to write the mass term for any type of perturbations on a cosmological background, 
\begin{equation}\label{general_mass_ term_FRW}
\mathcal{S}^{m}_{2}=-\frac{1}{2}\int d^4x\,\left[\mathcal{L}_{\gc\gc}^{(2)}+\mathcal{L}_{\fc\fc}^{(2)}+\mathcal{L}_{\gc\fc}^{(2)}+\mathcal{L}_{\gc\phi}^{(2)}+\mathcal{L}_{\fc\phi}^{(2)}+\mathcal{L}_{\phi\phi}^{(2)}\right],
\end{equation}
\small
\begin{align}
\mathcal{L}_{\gc\gc}^{(2)} & =\phi^{2}\left[\alpha_{\gc}h_{00}^2+\gamma_{\gc}h_{00}h_{ij}\delta^{ij}+2\epsilon_{\gc}h_{0i}h_{0j}\delta^{ij}+\eta_{\gc}h_{ij}h_{kl}\delta^{ij}\delta^{kl}+\frac{\sigma_{\gc}}{2}h_{ij}h_{kl}\left(\delta^{ik}\delta^{jl}+\delta^{il}\delta^{jk}\right)\right]\ \\
\hspace{-2 em}\mathcal{L}_{\fc\fc}^{(2)}& = \phi^{2}\left[\alpha_{\fc}\lc_{00}^2+\gamma_{\fc}\lc_{00}\lc_{ij}\delta^{ij}+2\epsilon_{\fc}\lc_{0i}\lc_{0j}\delta^{ij}+\eta_{\fc}\lc_{ij}\lc_{kl}\delta^{ij}\delta^{kl}+\frac{\sigma_{\fc}}{2}\lc_{ij}\lc_{kl}\left(\delta^{ik}\delta^{jl}+\delta^{il}\delta^{jk}\right)\right]\\\
\hspace{-2 em}\mathcal{L}_{\gc\fc}^{(2)}& = \phi^{2}\left[\alpha_{\gc\fc}h_{00}\lc_{00}+\gamma_{\fc\gc}h_{00}\lc_{ij}\delta^{ij}+\gamma_{\gc\fc}\lc_{00}h_{ij}\delta^{ij}+2\epsilon_{\gc\fc}h_{0i}\delta^{ij}\lc_{0j}+\eta_{\gc\fc}h_{ij}\lc_{kl}\delta^{ij}\delta^{kl}\right.\\
& +\left.\frac{\sigma_{gf}}{2}h_{ij}\lc_{kl}\left(\delta^{ik}\delta^{jl}+\delta^{il}\delta^{jk}\right)\right]\nn\\
\hspace{-2 em}\mathcal{L}_{\gc\phi}^{(2)}& = a^2\,\phi\,M_{\text{Pl}}\,\delta\phi\left[\zeta_g\,h_{00} +\xi_g\,h_{ij}\delta^{ij} \right]\\
\hspace{-2 em}\mathcal{L}_{\fc\phi}^{(2)}& = a^2\,\phi\,M_{\text{Pl}}\,\delta\phi\left[\zeta_f\,\lc_{00} +\xi_f\,\lc_{ij}\delta^{ij} \right]\\
\hspace{-2 em}\mathcal{L}_{\phi\phi}^{(2)}& = a^4\left[M_{\text{Pl}}^2\, \bar V+\frac{\partial^{(2)}W}{\partial\Phi\partial\Phi}\bigg|_{\bar g,\bar f} \right]\delta\phi^2\label{last}
\end{align}
\normalsize
The explicit form of the last term in eq. (\ref{last}) depends on the choice of the potential for the scalar field while $\bar{V}=V(\bar g, \bar f)$. 

We have at this point all the ingredients needed to study perturbations of the theory: it is just a matter of varying the total quadratic action (\ref{eqcovv}), where $\mathcal{S}^{kin}_{2}$ is the kinetic action evaluated on cosmological backgrounds and $\mathcal{S}^{m}_{2}$ is given by eq. (\ref{general_mass_ term_FRW}). We observe that taking the massive-gravity limit $\ell_{\mu\nu}\rightarrow 0$ in 
(\ref{general_mass_ term_FRW}), i.e. setting $\mathcal{L}_{\fc\fc}^{(2)}, \mathcal{L}_{\gc\fc}^{(2)}\,,\mathcal{L}_{\fc\phi}^{(2)}$ to zero in eq.  (\ref{general_mass_ term_FRW}), we get the generic parametrization of the mass term in a cosmological setting for a specific implementation of the ``mass-varying massive gravity'' model proposed in \cite{Huang:2012pe}, as discussed above. On the other hand, the results presented in \cite{Cusin:2015tmf} for the standard bi-gravity context are exactly recovered once the limit $\delta\phi\rightarrow 0$ is taken in (\ref{general_mass_ term_FRW}).

\textcolor{black}{As pointed out in \cite{Comelli:2014bqa}, in the context of standard bigravity, gradient exponential instabilities may arise in the scalar sector, therefore making the model not viable to describe the process of structure formation. In \cite{Konnig:2014xva} anyway it was shown that there exists a choice of parameters of the bigravity potential such that in the sub horizon limit, exponential gradient instabilities are absent in the scalar sector of perturbations. In this last work a model (the so-called $\beta_1-\beta_4$ model) was identified  to be the only one with both a viable background evolution and exponential gradient instabilities absent in the scalar sector. However, further investigations (see e.g \cite{Lagos:2014lca}, \cite{Cusin:2014psa}) pointed out that this sub-model suffers from another problem: in the scalar sector the Higuchi bound is violated during an early de\,Sitter inflationary phase, rendering it impossible to use the model for primordial cosmology, e.g. to embed the model in inflation.} 

\textcolor{black}{In our scale-free model, an additional scalar field is present, which at the perturbation level is mixing with the scalar perturbations of the metric. The mixing is only in the mass matrix (the kinetic structure is standard) of scalar perturbations. In principle, one would expect to find an analogous situation as before: for a special choice of parameters, gradient exponential instabilities are absent. Once a sub model (i.e. identified by the $\beta_i$ non vanishing) with such a good behaviour is pointed out, it would be interesting to consider the Higuchi bound for it. A full analysis of this type is quite involved and deserves a separate investigation, which we are planning to present in a future work.}

\section{A scale-free model of inflation and dark energy}\label{sec-inflation}

In this section we focus on the background evolution of scale-free bigravity in the cosmological ansatz of Sec. \ref{sec-FRW}. In particular we want to analyze if the model can be effectively used as a model of dark-energy at late times. Indeed, we know that  if the scalar field is non-dynamical (i.e. in the \emph{standard} bigravity scenario), in the cosmological ansatz, a phase of accelerated expansion can be recovered at late-times.
%with the role of dark energy played by the coupling with the second metric sector. 
The dark energy contribution becomes constant at late-times, when $\rho\rightarrow 0$, and it drives a quasi-de\,Sitter expansion phase.  In Sec. \ref{sec-inflation-2}, we will then turn to consider the case in which the scalar field is promoted to be the inflaton field and we study if it is possible to recover a viable inflationary scenario in this way. 
We focus on the second branch, which is the only one which can give rise to a viable cosmology.

  \subsection{Late-time accelerated expansion} \label{sec-inflation-1}

We start exploring which conditions need to be satisfied in order to get a late-time phase of accelerated expansion. For convenience we will here use cosmic time $t$ as the time variable ($dt\equiv a\,d\tau$) and indicate with  $\cdot$ derivatives with respect to cosmic time. We want the energy density $\rho_{\phi}$ defined in (\ref{e:rhophi}) to play the role of dark energy at late times. We underline that  in $\rho_{\phi}$ there is a contribution coming from the fact that we are dealing with a modified gravity model, and proportional to the bi-gravity potential $V(r, c)$ together with a contribution coming from the kinetic and potential terms of the scalar field\footnote{This last contribution is vanishing in the standard bigravity limit, i.e. $\Phi\rightarrow m$ and $W\rightarrow 0$.}. 

If we want accelerated expansion at late times to be driven by the (bigravity) mass term, we need (1) to choose the parameters of the model in such a way that $\rho_{\phi}$ gives the dominant contribution to the total energy  density at late times (2) to recover at late time a quasi de\,Sitter stage, i.e. $\rho_{\phi}\simeq -p_{\phi}$. 
\textcolor{black}{Note that in some sense this is an abuse of notation, since we really just want $\phi$ to freeze and let the massive interactions for the tensor (whose potential contributions are also captured by $\rho_{\phi},p_{\phi}$ as defined in \eqref{e:rhophi} and \eqref{state}) drive accelerated expansion in the same way as for standard bigravity.  Indeed, when $\phi$ is frozen, from an inspection of eq. (\ref{c}), we see that the expression defining the lapse $c$ in terms of the other background quantities is exactly the standard bigravity one. As a consequence, in this limit, eqs. (\ref{state2}) and (\ref{state22}), which give the energy density and pressure of the scalar field, have exactly the same form of the dark energy density and pressure in standard bigravity, with the addition of a constant contribution coming from the potential $W$.\footnote{In the limit of slowly evolving $\phi$, the contribution proportional to $W$ in eqs. (\ref{state2}) and (\ref{state22}) can be considered as renormalizing the $\beta_0$ term. }}

In this sense we can enforce the essence of our requirement via these conditions, which can be achieved choosing at some time $t_f$, $\phi(t_f)\sim H_0$ and 
\be\label{sr}
|\dot{\phi}(t_f)|\ll \phi(t_f)H(t_f)\,.
\ee
We observe that, assuming $\phi(t_f)\ll M_{\text{Pl}}$, this last condition implies 
\be\label{cons}
\dot{\phi}(t_f)^2\ll M^2_{\text{Pl}} H^2(t_f)\simeq |\mathcal{W}(\phi)|_{t_f} \,,
\ee
where in the last equality we have used the Friedmann equation for the $g$-metric. 
In the regime (\ref{sr}), the Bianchi constraint reduces to
\be
c\simeq \frac{\dot{r}+rH}{r H}=\frac{1}{H}\frac{\dot{b}}{b}\,,
\ee
which, once it is substituted in the Friedmann equation for the $f$-metric (\ref{F22}) gives
\be\label{yves}
3H^2=\frac{\phi^2}{2}\left(\frac{\beta_1}{r}+3\beta_2+3\beta_3 r+\beta_4 r^2\right)\,.
\ee
Substituting  in (\ref{yves}) $H$ from the Friedmann equation (\ref{F11}) (for $\rho\rightarrow 0$ and in the regime (\ref{sr})), we get the following condition 
\be
\frac{1}{M_{\text{Pl}}^2} W(\phi)+\frac{\phi^2}{2}\left(\beta_3 r^3+3\beta_2 r^2+3\beta_1 r+\beta_0-\frac{\beta_1}{r}-3\beta_2-3\beta_3 r-\beta_4 r^2\right)=0\,.
\ee
From this equation we read that if the condition (\ref{sr}) holds, the ratio between the two scale factors $r$ has to be constant, independently of the value of the $\beta_i$ and of the parameters of the potential $W$. Therefore, from the Friedmann equation (\ref{F11}), it follows  $H\simeq cnst$ and we recover a late-time de\,Sitter phase. In particular, since $c\simeq 1$, we get  an equation of state (\ref{eq-state}) for the fluid $p_{\phi} \sim - \rho_{\phi}$. Anyway, we observe that the condition (\ref{sr}) at late times can be satisfied only by fine-tuning the parameters of the model. Indeed, deriving  eq. (\ref{cons}) and substituting it in the equation of motion for the scalar field  (\ref{eqinf}),  we get 
\be\label{cu}
\frac{\dot{\phi}(t_f)}{H(t_f)\phi (t_f)}\simeq \frac{M_{\text{Pl}}^2}{H^2 (t_f)}\,\,V(r,c)\,|_{t_f}\,,
\ee
which is compatible with the slow-roll condition (\ref{sr}) only if the parameters of the bigravity potential are chosen in a such a way that $V(r, c)\sim H^2(t_f)/M^2_{\text{Pl}}$, i.e. $V$ is (severely) fine-tuned to be suppressed in this way. %{\bf Of course this solution is enormously fine-tuned.} %\footnote{We observe that the only way to render eq. (\ref{cu}) compatible with (\ref{sr}) is to choose the parameters of the bigravity potential in such a way that $V(r, c)\sim H^2(\tau_f)/M^2_{\text{Pl}}$. This solution is enormously fine-tuned.}.
It follows that using this model at late times as an effective model for dark energy requires a price to pay. The first option is to accept the above fine-tuning of the model parameters.\footnote{That such a fine-tuning is required is not surprising, since we are effectively using a massive scalar field, with mass $\simeq M_{\text{Pl}} V(r, c)$,  to drive late-time acceleration.}  Alternatively another possible way out is to give up our requirement on the model to be scale-free and introduce in the potential $W$ a constant contribution to play the role of dark energy at late times, thus essentially reintroducing an explicit cosmological constant. Needless to say that both options do not really present an improvement over the standard (cosmological constant problem plagued) $\Lambda$CDM solution.  Anyway, even if recovering viable dynamics at late times is not trivial in this model, if the scalar field is promoted to be the inflaton, this model can be used at early times as a model of inflation, with interesting phenomenological features, as we will explain in Sec. \ref{sec-inflation-2}.

\subsection{Early-time inflationary evolution}\label{sec-inflation-2}

This scale-free model of bigravity constitutes a generalization of the standard bigravity model, in which the mass parameter in front of the bigravity potential is promoted to a (dynamical) scalar field. The next step is to promote this scalar field to be the inflaton field and to study if it is possible to recover a viable inflationary scenario in this way. The remainder of this section is devoted to exploring this intriguing possibility. 

For definiteness, we specialize  to the case of a quartic potential for the inflaton field in eq. (\ref{W choice}). 
The field $\phi$ can in principle interact with other fields such as fermions, gauge bosons, etc., but we assume that this interaction can be neglected during inflation and that energy and pressure are dominated by the contribution from the inflaton. The energy-momentum tensor of $\phi$ is given by eq. (\ref{enmominflaton}). 
The effective potential defined in (\ref{effpotential0}) reads
\be\label{effsecond}
\mathcal{W}\left(\phi\right)=\frac{1}{2}\,m_{\phi}^2(\tau)\,\phi^2+\frac{\lambda}{4}\,\phi^4\,,\qquad m_{\phi}^2(\tau)\equiv M_{\text{Pl}}^2\, V(r,c)\,,
\ee
where we have defined a time-dependent mass for the inflaton and $V(r,c)$ is defined in eq. (\ref{mu}). Since the effective inflaton mass is a time-dependent quantity, the shape of the effective potential (\ref{effsecond}) changes with time. 

At early times we now impose the standard slow-roll condition on $\phi$ and we show that it is sufficient  to get an early de\,Sitter stage. We assume that at a given time $\tau_i$, there exists a region of space in which 
\be\label{sr2}
\phi'^2(\tau_i)\ll a^2 (\tau_i)\,\mathcal{W}(\phi)\,.
\ee
It follows that the Friedman equation (\ref{F12}) reduces to (considering $\rho\rightarrow 0$ at early time) 
\be
3 H^2\simeq 8\pi G\, \mathcal{W}(\phi)\,.
\ee
Using the slow-roll condition (\ref{sr2}), it follows 
\be
|\phi'(\tau_i)|\ll M_{\text{Pl}} \mathcal{H}(\tau_i)\apprle \phi(\tau_i)\,\mathcal{H}(\tau_i)\,,
\ee
where the last inequality follows from the standard assumption that at an early inflationary stage $\phi(\tau_i)\apprge M_{\text{Pl}}$. We therefore see that at early time the condition (\ref{sr}) which guarantees the existence of a de\,Sitter stage is automatically implemented as soon as the standard slow-roll conditions (\ref{sr2}) are imposed. Therefore, the same reasoning presented in section \ref{sec-inflation-1} applies and as a direct consequence of (\ref{sr2}) we obtain 
\be
r\simeq cnst\quad \rightarrow \quad c\simeq 1\quad \rightarrow \quad H\simeq cnst\,,
\ee
i.e. at early-time, a de\,Sitter-like inflationary stage is recovered.\footnote{The slow-roll condition (\ref{sr2}) used in the equation of motion for the inflaton leads to (\ref{cu}). This last equation at early times is consistent with the slow-roll condition assumed, for a proper choice of the model parameters (not necessarily fine-tuned in this case).}

%If we want to recover an usual slow-rolling phase at early times we need to choose in a proper way the parameters of the model. In particular, we need to chose the parameters of the model in such a way that at early times  $\phi'(\tau_i)\ll$\, $a^2W(\phi)$ and $r'\ll r \mathcal{H} $. From eq. (\ref{c}) it then follows $c\simeq 1$. These conditions guarantee that at early times we recover a quasi-de Sitter phase with $H\simeq const$, see eq. (\ref{F12}) with $\rho_{\phi}$ given by eq. (\ref{state2}). 

The evolution of the time-dependent mass depends on the details of the background evolution and in general it is different for different choices of the values of the parameters of the bigravity potential, $\beta_n$. We distinguish two different regimes. If we are in the regime in which $m_{\phi}^2(\tau)\geq 0$, the minimum of the effective potential is constant and given by $\phi=0$. If instead we have
$m_{\phi}^2(\tau)<0\,,$ 
then the shape of the effective potential is a double well and at a given instant of time $\tau$, the minimum is given by $\phi(\tau)=\pm m_{\phi}(\tau)/\sqrt{\lambda}$. 

In full generality, a transition between the two regimes is possible. If the evolution of $m_{\phi}^2$ is such that approaching the end of inflation it goes to positive values, then we recover the standard scenario, with the inflaton field oscillating around its constant (and vanishing) minimum configuration. After inflation, since then $\phi=0$, the coupling between the two metrics is vanishing and each of the two gravity sectors is evolving independently.  
 
The opposite situation, i.e. $m_{\phi}^2<0$ at late inflation, gives rise to a richer cosmology:  in this case the expectation value of the inflaton field is non vanishing and (in general) a time-dependent quantity. Therefore the cosmological dynamics of the background after inflation is more complicated, with the two gravity sectors interacting through the potential term. \textcolor{black}{This second scenario in the limit of $m_{\phi}\sim const$ corresponds to the so called dynamical branch in standard bigravity.}\footnote{\textcolor{black}{We observe that the condition $m_{\phi}\sim const$ after inflation is automatically realized if after inflation there exists a region of space in which the slow-roll condition (\ref{sr2}) is satisfied.}} After inflation the energy density of the inflaton defined in eq. (\ref{state2}) is still a time dependent quantity and plays the role of a dynamical dark energy contribution. However, we stress that the only way to get a \emph{positive} dark energy contribution at the end of inflation here, is to add to the inflaton potential $W$ a constant contribution, i.e. to introduce a cosmological constant-like term in the action, which also adds a new scale to the theory (and hence destroys its `scale-freeness').

\section{Conclusions}\label{sec-conclusions}

In this paper we have investigated a ``scale-free'' extension of massive (bi-)gravity models, where the mass parameter $m$ is promoted to be a dynamical scalar field. 
Our model is completely captured by the actions \eqref{vielbeinAction} and \eqref{metricAction} and its main features are the following:
\begin{itemize}
\item {\bf Strong coupling scales}: Perturbatively investigating the interaction structure of the theory around Minkowski, we find a strong coupling scale $\tilde\Lambda_3^3 = M_{\rm Pl}\phi_0^2$ in analogy with standard bi- and massive gravity (see e.g. equations \eqref{DL2} and \eqref{cubic-DL2}). Here we have employed a perturbative expansion of $\Phi = \phi_0 +\delta\phi$ around a fixed and non-dynamical reference value $\phi_0$. Different to the standard massive (bi-)gravity cases, enforcing tadpole cancellation conditions is crucial for this strong coupling scale to be made explicit.
\item {\bf Modified low-energy physics}: Additional interactions, not present in standard massive and bi-gravity decoupling limits, can be found in our model due to the new dynamical scalar degree of freedom $\Phi$. They generically affect scaling limits and the low-energy physics of the model. We capture these new interactions in a scaling limit (see e.g. equations \eqref{DL1},\eqref{cubic-DL1} and \eqref{quartic-DL1}) different from the standard decoupling one, which clearly illustrates the regimes where our pertubative expansion is valid and the conditions necessary to satisfy it.
\item {\bf Cosmological framework}: Exploiting the method presented in \cite{Cusin:2015tmf}, in section \ref{sec-Quad} we derived the quadratic action of our theory around both generic and explicitly FLRW background configurations. We find the precise form of new interactions due to the additional scalar $\Phi$ complementing the standard bigravity ones derived in \cite{Cusin:2015tmf}. This will also enable the detailed study of cosmological perturbations in future work. Furthermore we have derived the background dynamics and established the nature of the two different branches of solutions in our model.
\item {\bf Dark Energy and Inflation}: Finally we explicitly study how periods of early and late-time acceleration, i.e. an inflationary and dark energy phase, can be realised at the background level in our model. Inflation can be successfully (and without fine-tuning parameters) realized, with the scalar field $\Phi$ acting as a slowly-rolling inflaton and leading to an early-time quasi-de\,Sitter inflationary stage (see section \ref{sec-inflation-2}). Surprisingly, generating a period of late-time acceleration is significantly more difficult and requires either extreme fine-tuning of parameters or the (re-)introduction of an explicit additional (mass) scale in the potential for $\Phi$ (see section \ref{sec-inflation-1}).\footnote{More specifically, if we want the inflaton energy density to play the role of a (positive) dark energy contribution at late times, the inflaton after inflation has to sit on a positive minimum of its effective potential. This can be realized only  by introducing a positive shift in the inflaton potential (i.e. a cosmological constant like term in the action), to play the role of dark energy at late times.} This restriction will also apply to our massive gravity limit and generic mass-varying models such as \cite{Huang:2012pe}.
\end{itemize}
Throughout this paper we have considered both a scale-free extension of bigravity and also its massive gravity limit. Various extensions are worthy of further investigation, ranging from extending the work presented here to fully-fledged multi-gravity models (see appendix \ref{appendix-MG}) to considering couplings of matter and/or the additional scalar $\Phi$ to more than one metric (i.e. going beyond the minimal coupling of GR). 
The perturbative properties of the model in a cosmological setting as well as a study of the explicit evolution of strong coupling scales in different background configurations and throughout different phases are also left for future work. 
Finally, and in the spirit of scale-freedom, it would be interesting to embed our approach in a fully scale-free framework where the Planck mass(es) are also promoted to become dynamical (scalar) fields and their present-day fixed nature arises via spontaneous symmetry breaking along the lines of \cite{Shaposhnikov:2008xb,GarciaBellido:2011de,Kannike:2015apa,Einhorn:2015lzy,Einhorn:2016mws,Ferreira:2016vsc}. 

We conclude by summarising and emphasising the defining features of the scale-free extension to massive (bi-)gravity considered here. This extension eliminates one of the mass scales in the original theory, replacing it by a dynamical field and in the process can alleviate low strong-coupling scale problems \textcolor{black}{at early times which hinder the predictivity of the theory then, essentially via having $\Phi \gg H_0$ at early times whilst $\Phi$ eventually transitions towards smaller values at late times. This does enable us to nicely describe inflationary physics within this model, although a successful period of late-time acceleration (i.e. obtaining $\Phi \sim H_0$ at late times)} requires resorting to either fine-tuning or the (re-)introduction of a separate mass scale in the potential for $\Phi$. We hope that our work both helps to clarify the nature of scale-freeness in and mass-varying extensions of massive bigravity and paves the way to understand and fully extract the physical signatures of these models.
\\

\noindent {\bf Acknowledgements:} We would like to thank Ruth Durrer and Filippo Vernizzi for an interesting discussion which led to this investigation. The work of GC is supported by the Swiss National Science Foundation. JN acknowledges support from the Royal Commission for the Exhibition of 1851 and BIPAC.

\newpage

\appendix

\section{Scale-free multi-gravity} \label{appendix-MG}

The vielbein version of our extended bigravity theory \eqref{vielbeinAction} can be straightforwardly extended to analogous ``scale-free'' (up to the Planck masses for the different fields) multi-gravity theories. The generalised ghost-free multi-gravity theory reads
\begin{align}
{\cal S}_{\rm MG} &=  \sum_{(i)}^N \frac{M_{\rm Pl}^2}{4} \int \epsilon_{ABCD} {\bf E}^A_{(i)} \wedge {\bf E}^B_{(i)} \wedge {\bf R}^{CD}\left[E_{(i)} \right] -\nn \\ &- \frac{M_{\rm Pl}^2}{2} \sum_{(i,j,k,l)}^N  \beta_{(i,j,k,l)} \int \; \Phi^2 \,\epsilon_{ABCD} \;  {\bf E}^A_{(i)} \wedge {\bf E}^B_{(j)}\wedge {\bf E}^C_{(k)}\wedge {\bf E}^D_{(l)} +\nn \\
&+ \int d^4 x \det {\tilde E}\Big( \frac{1}{2}\tilde \nabla_\mu  \Phi \tilde \nabla^\mu \Phi - W(\Phi) \Big)  + \int d^4x \; \det \tilde E \; {\cal L}_m \left[\tilde E,\Psi_i\right],
\label{MultiAction}
\end{align}
where we have set all Planck masses to be identical as before (this is straightforwardly generalisable), the $\beta$ parameters are dimensionless constant coefficients completely symmetric in their label indices $(i,j,k,l)$ and we have allowed matter and the scalar $\Phi$ to minimally couple to the general effective vielbein construction for consistent matter couplings \cite{deRham:2014naa,Noller:2014sta,Melville:2015dba}
\be
\tilde E = \sum_i \alpha_{(i)} E_{(i)}\,,
\ee 
where the $\alpha_{(i)}$ are dimensionless constant coefficients. As before each spin-2 field comes equipped with an Einstein-Hilbert term (first line in \eqref{MultiAction}), we have massive (multi-)gravity interactions with a graviton mass that has been promoted to be a field $\Phi$ (second line), and we have an additional piece of the action, giving $\Phi$ dynamics, as well as a coupling of gravity to matter fields $\Psi_i$ (third line). Note  that we have chosen to write explicit expressions for four space-time dimensions -- generalisation to arbitrary $D$ dimensions is straightforward.

\section{The perturbed mass term}\label{perturbed mass}

In this appendix, we present additional details for derivation of the quadratic mass term of scale-free bigravity around generic backgrounds, discussed in Sec. \ref{sec-Quad}.
Using the technique described in \cite{Cusin:2015tmf}, it is possible to write the perturbations for the object $\mathbb{X}=\sqrt{g^{-1} f}$ in terms of perturbations of $g^{-1} f=g^{\mu\rho}f_{\nu\rho}$.\footnote{We omit here the detailed description of the procedure, which is the main core of \cite{Cusin:2015tmf}.} In writing the perturbed mass term, we keep only terms up to second order in the metric and scalar field perturbations. Therefore, for example, the fundamental quantity $g^{-1}f$ is expanded as
\be
g^{\mu\rho}f_{\rho\nu}=(\delta^{\mu}_{\alpha}-\mathbb{h}^{\mu}_{\alpha}+\mathbb{h}^{\mu}_{\gamma}\mathbb{h}^{\gamma}_{\alpha})\,\bar g^{\alpha\rho}\bar f_{\rho\beta}\,(\delta^{\beta}_{\nu}+\mathbb{l}^{\beta}_{\nu}) + \mathcal{O}(\mathbb{h}^3)\,.
\ee

Up to second order in the perturbations $\mathbb{h}_{\mu\nu}$ and $\mathbb{\lc}_{\mu\nu}$, the bigravity potential can be written as\footnote{We will always denote the indices of $h$ with the letters $\mu\nu$ and the indices of $\lc$ with the letters $\al\beta$ in the mixed term, $\MM^{\mu\nu\al\beta}_{\gc\fc}(\bar f,\bar g)h_{\mu\nu}\lc_{\al\beta}$.}
\begin{align}
\sqrt{-\det g}\,\,V(f,g) =& \sqrt{-\bar g}\,\Big[V(\bar f,\bar g)  +\MM^{\mu\nu}_{\gc}(\bar f, \bar g)\mathbb{h}_{\mu\nu} +\MM^{\mu\nu}_{\fc}(\bar f, \bar g)\mathbb{l}_{\mu\nu} +   \\ 
 &+ \MM^{\mu\nu\al\beta}_{\gc\gc}(\bar f,\bar g)\mathbb{h}_{\mu\nu}\mathbb{h}_{\al\beta} + \MM^{\mu\nu\al\beta}_{\gc\fc}(\bar f,\bar g)\mathbb{h}_{\mu\nu}\mathbb{l}_{\al\beta} + \MM^{\mu\nu\al\beta}_{\fc\fc}(\bar f,\bar g)\mathbb{l}_{\mu\nu}\mathbb{l}_{\al\beta}\Big] \,,\nn
\end{align}
where\small
\begin{align} \label{e:Mmunu0}
\MM^{\mu\nu}_{\gc}(\bar f,\bar g) &\equiv  \frac{1}{\sqrt{- g}}\frac{\partial(\sqrt{- g}\,\,V(f,g))}{\partial g_{\mu\nu}}\bigg|_{g=\bar{g}, f=\bar{f}}\,, \\
\MM^{\mu\nu}_{\fc}(\bar f,\bar g) &\equiv \frac{1}{\sqrt{- g}}\frac{\partial(\sqrt{-g}\,\, V(f,g))}{\partial f_{\mu\nu}}\bigg|_{g=\bar{g}, f=\bar{f}}\,,\\ 
\MM^{\mu\nu\al\beta}_{\gc\gc}(\bar f,\bar g) &\equiv  \frac{1}{2}\frac{1}{\sqrt{-g}}\frac{\partial^{2}(\sqrt{-g}\,\, V(f,g))}{\partial g_{\mu\nu}\partial g_{\al\beta}}\bigg|_{g=\bar{g}, f=\bar{f}}\,, \\
\MM^{\mu\nu\al\beta}_{\gc\fc}(\bar f,\bar g) &\equiv  \frac{1}{\sqrt{-g}}\frac{\partial^{2}(\sqrt{- g}\,\, V(f,g))}{\partial g_{\mu\nu}\partial f_{\al\beta}}\bigg|_{g=\bar{g}, f=\bar{f}}\,, \\
\MM^{\mu\nu\al\beta}_{\fc\fc}(\bar f,\bar g) &\equiv  \frac{1}{2}\frac{1}{\sqrt{-g}}\frac{\partial^{2}(\sqrt{-g} \,\,V(f,g))}{\partial f_{\mu\nu}\partial f_{\al\beta}}\bigg|_{g=\bar{g}, f=\bar{f}}\,.\label{e:Mmunuf}
\end{align}\normalsize
The mass matrices $\MM_{\bullet\bullet}^{\mu\nu\al\beta}(\bar{g}\,,\bar{f})$ have been calculated for the first time in \cite{Cusin:2015tmf} in the context of standard bigravity. For mass-varying bigravity, the expression of $\MM_{\bullet\bullet}^{\mu\nu\al\beta}(\bar{g}\,,\bar{f})$ in terms of background matrices are exactly those of \cite{Cusin:2015tmf} with the replacement $m^2\rightarrow 1/2$. The terms linear in the metric perturbations $\MM_{\bullet}^{\mu\nu}$ in the standard bigravity formulation of  \cite{Cusin:2015tmf} were canceling on shell. In our model they give genuinely new mass terms, mixing $\Phi$ perturbation and metric perturbations. These are explicitly given by \small
%\textcolor{red}{GC: Check the overall sign in the expressions below. This is very important}
\begin{align}
&\mathcal{M}^{\mu\nu}_{g}(\bar f, \bar g)\equiv \frac{\delta V}{\delta g_{\mu\nu}}+\frac{1}{2}g^{\mu\nu}\,V\bigg|_{\bar{g},\bar{f}}=-\frac{1}{4}\, \sum_{n=0}^3(-)^{n+1}\, \beta_n\, \left[\bar g^{\mu\lambda}\, Y^{\nu}_{(n)\lambda}\left(\mathbb X_g\right)+\bar g^{\nu\lambda}\, Y^{\mu}_{(n)\lambda}\left(\mathbb X_g\right)\right]\,,\\
&\mathcal{M}^{\mu\nu}_{f}(\bar f, \bar g)\equiv \frac{\delta V}{\delta f_{\mu\nu}}+\frac{1}{2}f^{\mu\nu}\,V\bigg|_{\bar{g}, \bar{f}}= -\frac{1}{4}\, \sum_{n=0}^3(-)^{n+1}\, \beta_{4-n}\,\left[\bar f^{\mu\lambda}\, Y^{\nu}_{(n)\lambda}\left(\mathbb X_f\right)+\bar f^{\nu\lambda}\, Y^{\mu}_{(n)\lambda}\left(\mathbb X_f\right)\right]\,,
\end{align}
\normalsize 
where $\mathbb X_g=\sqrt{\bar{g}^{-1} \bar{f}}$\,, $\mathbb X_f=\sqrt{\bar{f}^{-1} \bar{g}}$. 

The quadratic kinetic action for gravity (both for $g$ and $f$) can be written in term of the Lichnerovitz operators on curve space time as in eq. (\ref{eqkin}), where 
\begin{align}
\EE^{\mu\nu\alpha\beta}(\bar g) =&\frac{1}{4}\Big[\left(\bar g^{\mu\alpha}\bar g^{\nu\beta}-\bar g^{\mu\nu}\bar g^{\alpha\beta}\right)\Box+\left(\bar g^{\mu\nu}\bar g^{\alpha\rho}\bar g^{\beta\sigma}
+ \bar g^{\alpha\beta}\bar g^{\mu\rho}\bar g^{\nu\sigma} - \bar g^{\mu\beta}\bar g^{\nu\rho}\bar g^{\alpha\sigma}- \bar g^{\alpha\nu}\bar g^{\beta\rho}\bar g^{\mu\sigma}\right)\nabla_\rho\nabla_\sigma\Big] +\nn\\
&-\frac{R(\bar{g})}{8}\left(\bar{g}^{\mu\alpha}\bar{g}^{\nu\beta}+\bar{g}^{\nu\alpha}\bar{g}^{\mu\beta}-\bar{g}^{\mu\nu}\bar{g}^{\alpha\beta}\right)-\frac{1}{4}\left(\bar{g}^{\mu\nu}\,R^{\alpha\beta}(\bar{g})+\bar{g}^{\alpha\beta}\,R^{\mu\nu}(\bar{g})\right)+\nn\\
&+\frac{1}{4}\left(\bar{g}^{\mu\al}R^{\beta\nu}(\bar{g})+\bar{g}^{\mu\beta}R^{\alpha\nu}(\bar{g})+\bar{g}^{\nu\al}R^{\beta\mu}(\bar{g})+\bar{g}^{\nu\beta}R^{\alpha \mu}(\bar{g})\right) \,,
\label{e:EEg}
\end{align} 
and an analogous result holds for $f$.

\section{Parametrization of the cosmological mass term}\label{FRW}
 
We give here the explicit expressions for the functions which parametrize the mass tensor on cosmological  backgrounds, as presented in Section \ref{massesF}: 
 \small

\begin{eqnarray} 
\alpha_{\gc} & = & -\frac{1}{4}\left(\beta_{0}+\beta_{3}r^{3}+3\beta_{2}r^{2}+3\beta_{1}r\right),\label{ah} \\
\gamma_{\gc} & = & -\frac{1}{4}\left(\beta_{0}+\beta_{2}r^{2}+2\beta_{1}r\right),\\
\epsilon_{\gc} & = & \frac{1}{4}\left(\beta_{0}+\frac{(3c+2)r}{c+1}\beta_{1}+\beta_{2}\frac{(3c+1)r^{2}}{c+1}+\frac{\beta_{3}cr^{3}}{c+1}\right),\\
\eta_{\gc} & = & \frac{1}{4}\left(\beta_{0}+\beta_{2}cr^{2}+\beta_{1}(c+1)r\right),\\
\sigma_{\gc} & = & -\frac{1}{4}\left(2\beta_{0}+\beta_{3}cr^{3}+\beta_{2}(3c+1)r^{2}+\beta_{1}(2c+3)r\right),\\
\nonumber\\
\alpha_{\fc} & = & -\frac{1}{4c^{3}}\left(\beta_{4}+3\beta_{3}r+3\beta_{2}r^{-2}+\beta_{1}r^{-3}\right),\\
\gamma_{\fc} & = & -\frac{1}{4c}\left(\beta_{4}+2\beta_{3}r^{-1}+\beta_{2}r^{-2}\right),\\
\epsilon_{\fc} & = & \frac{1}{4c}\left(\beta_{4}+\frac{\beta_{3}(2c+3)r^{-1}}{(c+1)}+\frac{\beta_{2}(c+3)r^{-2}}{(c+1)}+\frac{\beta_{1}r^{-3}}{(c+1)}\right),\\
\eta_{\fc} & = & \frac{1}{4}\left(\beta_{4}c+\beta_{3}(c+1)r^{-1}+\beta_{2}r^{-2}\right),\\
\sigma_{\fc} & = & -\frac{1}{4}\left(2\beta_{4}c+\beta_{3}(3c+2)r^{-1}+\beta_{2}(c+3)r^{-2}+\beta_{1}r^{-3}\right),\\
\nonumber \\
\alpha_{\gc\fc} & = & 0,\\
\gamma_{\gc\fc} & = & -\frac{1}{2r}\left(\beta_{1}+2\beta_{2}r+\beta_{3}r^{2}\right),\\
\gamma_{\fc\gc} & = & -\frac{1}{2rc}\left(\beta_{1}+2\beta_{2}r+\beta_{3}r^{2}\right),\\
\epsilon_{\gc\fc} & = & \frac{1}{2(1+c)r}\left(\beta_{1}+2\beta_{2}r+\beta_{3}r^{2}\right),\\
\eta_{\gc\fc} & = & \frac{1}{2r}\left(\beta_{1}+\beta_{3}cr^{2}+\beta_{2}(c+1)r\right),\\
\sigma_{\gc\fc} & = & -\frac{1}{2r}\left(\beta_{1}+\beta_{3}cr^{2}+\beta_{2}(c+1)r\right),\label{shl}
\end{eqnarray}
\begin{eqnarray} 
\zeta_{\gc}&=& \left(\beta_0+3\beta_1r+3\beta_2 r^2+\beta_3r^3\right),\\
\xi_{\gc}&=&-\left(\beta_2+\beta_1(2+c)r+\beta_2(1+2c)r^2+\beta_3 c r^3\right),\\
\zeta_{\fc}&=&\frac{1}{r^2 c^2} \left(\beta_4+3\frac{\beta_3}{r}+3\frac{\beta_2}{r^2}+\frac{\beta_1}{r^3}\right),\\
\xi_{\fc}&=&-\frac{1}{r^2 c^2}\left(c\beta_4+\frac{\beta_3}{r}(2c+1)+\frac{\beta_2}{r^2}(2+c)+\frac{\beta_1}{r^3} \right),
\end{eqnarray}
\begin{eqnarray} 
 \bar{V}&=& \beta_0+\beta_1\left(c+3\right) r+3\, \beta_2 (c+1) r^2+\beta_3 \left(1+3c\right)r^3+\beta_4\, c \,r^4.
\end{eqnarray}

\newpage 
%\subsection*{Acknowledgments}

\newpage

\bibliographystyle{JHEP}
\bibliography{biphi}

\end{document}